\documentclass{ptptex}

\usepackage{hyperref}
\usepackage{epic,amsmath,graphicx,wrapft}

\def\mb#1         {\mbox{\boldmath $#1$}}

\title{Construction of Lorentz-Invariant Amplitudes
from Rest-Frame \\ Wave Functions 
in HQET \\
--  {\it Application to the Isgur-Wise Function}  --}

\notypesetlogo
\author{
Takayuki \textsc{Matsuki},$^{1,}$\footnote{E-mail: matsuki@tokyo-kasei.ac.jp} and 
Kohichi \textsc{Seo}$^{2,}$\footnote{E-mail: seo@gifu-cwc.ac.jp}
}

\inst{
$^1$Tokyo Kasei University, Tokyo 173-8602, Japan \\
$^2$Gifu City Women's College, Gifu 501-0192, Japan \\
}

\recdate{May 28, 2007}%            %Editorial Office will fill in this.

\abst{
Succeeding in predicting $0^+$ and $1^+$ states of $D$ and $D_s$ heavy mesons with
our semi-relativistic quark potential model, we examine a method for constructing
Lorentz-invariant scattering amplitudes and/or decay widths and develop a formulation
to calculate Lorentz-boosted ones given the rest-frame wave functions in our model.

To illustrate the effectiveness of our method, we apply the formulation to the problem
of calculating the
semileptonic weak form factors from the rest-frame wave functions of heavy
mesons and numerically calculate the dynamical $1/m_Q$ corrections to those for
the process $\bar B \rightarrow D^{(*)}\ell\bar\nu$ based on our model of heavy mesons.
It is shown that nonvanishing expressions for
$\rho_1(\omega)=\rho_2(\omega)$ and $\rho_3(\omega)=\rho_4(\omega)=0$ are obtained
in a special Lorentz frame, where $\rho_i(\omega)$ are the parameters used in the
Neubert-Rieckert decomposition of form factors.
Various values of form factors are estimated. These values are compatible with recent
experimental data as well as other theoretical calculations.
}

\begin{document}
\maketitle  %% ptptex

\section{Introduction}
\label{sec:introduction}
The discovery of the narrow meson states $D_{s0}(2317)$ by BaBar \cite{BaBar03}
and $D_{s1}(2460)$ by CLEO \cite{CLEO03} and the following confirmation by
Belle \cite{Belle03,FOCUS04} has motivated many theorists to explain these states, as
previous studies of these states (Refs. \citen{GIK} and \citen{EGF98})
using quark potential model apparently fail to reproduce these mass values.
More recent experiments have found many other heavy-light mesons : broad
$D_{0}^{*}(2308)$ and $D_{1}'(2427)$ mesons found by the
Belle collaboration \cite{Belle04}, which are identified as $c\bar q$ ($q=u/d$)
excited ($\ell=1$) bound states and have the same quantum numbers, $j^P=0^+$ and
$1^+$, as $D_{sJ}$, respectively (The masses of these mesons have been reported with quite
different values in the CLEO and FOCUS experiments, and hence are not yet determined with
definiteness. \cite{CLEO00,FOCUS04});
narrow $B$ and $B_s$ states of $\ell=1$, $B_1(5720)$, $B_2^*(5745)$, and $B_{s2}^*(5839)$
found by CDF and D0, \cite{CDF_D006} whose decay
widths are also narrow because of their decay through the D-waves; and seemingly radial
excitations $(n=2)$ of the $0^+$ $D_{s0}(2860)$ state found by BaBar \cite{Palano06}
and the $1^-$ $Ds^*(2715)$ state found by Belle \cite{Abe06}.
Furthermore several $c\bar c$ quarkonium-like states have been discovered :
$X(3872)$, $X(3940)$, $Y(3940)$, $Z(3930)$, and $Y(4260)$.

The discovery of these mesons triggered a series of studies of the spectroscopy of heavy-light
and/or heavy-heavy hadrons using many kinds of ideas, including our semirelativistic
quark patential model. Because we do not discuss these ideas in this paper,
we refer the reader to the review articles \cite{Swanson06,Rosner06,Colangelo04}.
In previous
papers \cite{Matsuki97}, we formulated a method for calculating the spectrum of
heavy-light mesons in order to construct the Schr\"odinger equation for a bound state.
In that method, the
mass is expressed as an eigenvalue of the Hamiltonian of the heavy-light system, in
which a bound state consists of a heavy quark and a light antiquark, and negative
energy states of a heavy quark appear in the intermediate states when calculating an
energy eigenvalue.
In oder to make our formulation more reliable, we attempt to reproduce narrow $B$ and $B_s$
states of $\ell=1$, $B_1(5720)$, $B_2^*(5745)$ and $B_{s2}^*(5839)$, together with higher
states of $J^P$ for $D$, $D_s$, $B$ and $B_s$ \cite{Matsuki05}, and also to reproduce
radial excitations of the $0^+$ state of $D_{s0}(2860)$ and the $1^-$ state of $Ds^*(2715)$
\cite{Matsuki06}. These calculations show our model's reliability, as these results
fit well with experiments.
What we need to do next is to show that our approach can also produce a method for calculating
scattering amplitudes and decay widths using the rest-frame wave functions. Here, we
concentrate on the construction of such a formulation, and therefore we refer readers the
review paper Ref. \citen{Colangelo06} for explanations of other methods and their results,
because there are many papers that give estimates of widths of many decay modes of heavy mesons.

The heavy quark effective theory (HQET) is a very convenient tool
for studying heavy mesons/baryons, including at least one heavy 
quark.\cite{HQET,REVIEW} In particular, there are many studies on the 
Isgur-Wise function, \cite{POL}\tocite{CHENG04} but most of the
calculations in those works use a simple-minded Gaussian form for the meson wave 
function or a solution to a single-particle Dirac equation in a potential.
Some other people have used the Bethe-Salpeter solution for a meson wave 
function, but it seems there has been no systematically developed way to
calculate the higher-order terms in $1/m_Q$, \cite{KUGO,HUANG} although some
papers \cite{LIU} claim to have done so.
Other people \cite{CHENG04} used the light-front formalism to calculate the scattering
amplitudes, so that in the heavy quark limit they obtained the Isgur-Wise function,
in which they adopted the Gaussian-type wave function for the heavy mesons.
There is a recent paper \cite{EGF06}
in which the authors use a "relativistic" formulation to calculate the masses
of heavy mesons and apply it to calculate the Isgur-Wise functions for
semileptonic $B$ decays. Although that paper adopts an approach that is different from
ours, they present a method for calculating higher-order corrections in $1/m_Q$ to form
factors, which is one of the purposes of this paper.

Most of the other calculations of form factors used the so-called trace
formalism developed by Falk et al., \cite{HQET} which assumes the Lorentz covariance
of the final form and fits it to the experimental data, or the QCD sum rule
\cite{Neubert1} was used to obtain the relation with the quark/gluon condensates.
In previous papers \cite{Matsuki97}, we developed
a semi-relativistic formulation for calculating the meson masses and wave 
functions for the heavy-light $Q\bar q$ system with one heavy quark, $Q$, and one light
antiquark, $\bar q$,
introducing phenomenological dynamics. That is, assuming a Coulomb-like vector potential
as well as a confining scalar potential to $Q\bar q$ bound states, we expanded 
the effective Hamiltonian and then perturbatively solved the Schr\"odinger equation 
as an expansion in $1/m_Q$.
The meson wave functions thereby obtained and expanded in $1/m_Q$ can be used in principle
to calculate ordinary form factors or Isgur-Wise functions and their corrections 
higher-order in $1/m_Q$ for semileptonic weak or other decay processes.
However, what we have obtained are wave functions in the rest-frame, and therefore we need
to develop a method to obtain Lorentz-invariant amplitudes or Lorentz-boosted wave
functions. In this paper we would like to address the problem of how accurately we can
calculate scattering amplitudes and/or decay widths of heavy-light systems using
the rest-frame wave functions.
In the following, explaining the results obtained in Ref. \citen{Matsuki97}, we show how
to calculate the Lorentz-invariant amplitudes by taking the Isgur-Wise function as
an example. This calculation is general enough to apply our formulation to other scattering/decay
amplitudes.

Following a previous study, \cite{Seo97} in which we derived the zeroth-order
form of the Isgur-Wise function expanded in $1/m_Q$, we develop a
formulation for calculating higher-order corrections to the semi-leptonic weak form
factors in $1/m_Q$ from our semi-relativistic wave functions for heavy mesons in the
rest-frame obtained in Ref. \citen{Matsuki97}.
The problem involved in constructing a Lorentz-boosted wave function is that there is
ambiguity in determining the space-time coordinates of two composite particles from
information regarding one bound state, in this case a heavy meson. We study four
cases for the reference frames of composite particles and then give a prescription
for calculating matrix elements of currents using the rest-frame wave functions in
\S \ref{sec:formulation}.
In \S \ref{sec:IW}, we give the form factors at zeroth order for four
reference frames, show that they agree with each other in the HQET limit, and then obtain
them at first-order for one special reference frame. The
calculation to lowest order in $1/m_Q$ gives the numerical value of the slope for 
the Isgur-Wise function at the origin, and the semi-leptonic weak form factors
are calculated up to first-order in $1/m_Q$ in \S \ref{sec:Vcb}.
It is found that there are no dynamical contributions to the form factors;
i.e., the first-order corrections to the wave functions do not contribute to
the form factors.
In \S \ref{sec:Vcb}, studying $\bar B \rightarrow D\ell\nu$ and
$\bar B \rightarrow D^{*}\ell\nu$ processes, physical quantities related to the
CKM matrix elements are obtained.
Finally, \S \ref{sec:summary} is devoted to summary and discussions of the results obtained
in this paper.
All the details associated with calculations of the form factors are given in the appendices.

\section{Formulation}
\label{sec:formulation}
\subsection{Schr\"odinger equation}
\label{sub:FYeq}
In Ref. \citen{Matsuki97}, we calculated the
mass spectrum of heavy mesons in a so-called Cornell potential model, 
which includes both scalar and vector potentials.
Although we needed only a stationary system, or rest-frame wave
function there, we must now treat a moving frame of a heavy 
meson to calculate weak form factors. Although we cannot work in a fully
relativistic system as far as the Hamiltonian formulation is concerned, 
we adopt the notation of the Nambu-Bethe-Salpeter equation as much as 
possible. First, we define a heavy meson ($X$) wave function composed of a 
light anti-quark, $q^c(x)$, and a heavy quark, $Q(y)$, as
\begin{equation}
  \left<0\right|q_\alpha^c(t,\vec x)\,Q_\beta(t,\vec y)
  \left|X;P_X\,\{\ell\}\right>={\psi_X^{\ell}}_{\alpha\,\beta}
  \left(\left(0,\vec x-\vec y\right); P_X\right)\,e^{-iP_X\cdot y},
\end{equation}
where $\alpha$ and $\beta$ are Dirac indices of four-spinors,
${P_X}_\mu$ is the heavy meson four-momentum, $\{\ell\}$ is a set of quantum
numbers, and ${P_X}\cdot y={P_X}_\mu\, y^\mu$. Note that this definition of
the wave function, $\psi_X^\ell$, equals to that used in Ref. \citen{Matsuki97}
multiplied by -1. The heavy meson state is an eigenstate of the four-momentum 
operator ${\cal P_\mu}$ given by
\begin{equation}
  {\cal P_\mu} \left|X;P_X\,\{\ell\}\right>
  ={P_X}_\mu \left|X;P_X\,\{\ell\}\right>, 
  \qquad {P_X}_0=\sqrt{M_X^2 + \vec P_X^2},
\end{equation}
with $M_X$ a heavy meson mass. In this case, the Schr\"odinger equation
in the $X$ moving frame is given by
\begin{equation}
  H\,\psi_X^{\ell}\left(\left(0,\vec x-\vec y\right);P_X\right) =
  \sqrt{M_X^2+\vec P_X^2}\,\psi_X^{\ell}\left(\left(0,\vec x-\vec y\right);
  P_X\right),\label{eq:FYeq}
\end{equation}
where the Hamiltonian is given by
\begin{equation}
  H=({\vec \alpha}_q\cdot {\vec p} +\beta_q m_q)
  +\left[{\vec \alpha}_Q\cdot ({\vec P_X}-{\vec p}) +\beta_Q m_Q\right]
  +\sum_{i,\,j}\beta_q O_{q\,i} V_{i\,j}({\vec x}-{\vec y})\beta_Q O_{Q\,j}.
  \label{eq:hamil1}
\end{equation}
Here, $\vec p=-i{\vec \nabla}_x$ and $\vec P_X$ is the heavy meson momentum. 
In our case, Eq.~(\ref{eq:hamil1}) gives $O_{q\,i}=O_{Q\,i}=1$
for the scalar potential, $V_{i\,j}(\vec x -\vec y)=S(r)$ and $O_{q\,i}
={\gamma_q}_\mu$ and $O_{Q\,j}={\gamma_Q}^\mu$ for the vector potential, 
$V_{i\,j}(\vec x -\vec y)=V(r)$ with $r=\left|\vec x -\vec y\right|$. 
Actually, we have included the transverse part of the vector potential and the 
total potential in Eq.~(\ref{eq:hamil1}) is given by 
\begin{equation}
  \beta _q\beta _Q\,S(r)+\left\{ {1-\frac{1}{2}\left[\, {\vec \alpha _q\cdot
  \vec \alpha _Q}+\left( {\vec \alpha _q\cdot \vec n} \right)
  \left( {\vec \alpha_Q\cdot \vec n} \right)\, \right]} \right\}V(r), 
  \label{eq:pot2}
\end{equation}
with
\begin{equation}
  S(r)=\frac{r}{a^2} + b, \qquad
  V(r)=-\frac{4}{3} \frac{\alpha_s}{r}, \qquad
  \vec n=\frac{\vec r}{r}.\label{eq:pot}
\end{equation}
In Ref. \citen{Matsuki97}, we solved the 
Schr\"odinger equation given by Eq.~(\ref{eq:FYeq}) with $\vec P_X=\vec 0$ 
order by order in $1/m_Q$ by expanding $H$, $\psi_X^\ell$, and $M_X$ 
in $1/m_Q$, in order to numerically calculate the mass spectrum, i.e., an eigenvalue,
$M_X$, in each order of perturbation, and then to relate the results to 
those of HQET.

\subsection{Lorentz boost}
\label{sub:lorentz}
In order to calculate the decay rate, we need to Lorentz boost the rest-frame
wave function that we obtained in previous papers. 
To explain the idea, we consider the ${\bar B}\rightarrow D^{(*)}\ell\nu$ decay 
process as an example and express the wave function in the moving frame in 
terms of that in the rest-frame.

Assume that the heavy meson is moving in the $+z$ direction with a velocity 
$\beta$. Below, we consider two cases of composite particles, i.e.,
$q^c$ and $Q$. In the following, for the initial state of a heavy meson, we adopt the 
notation $q^c(x^0,\vec x)$ and $Q(y^0,\vec y)$ and for the final
state, $q^c(x'^0,\vec x\,')$ and $Q(y'^0,\vec y\,')$. The relation
between the constituent particles in the rest and moving frames is given by
\begin{eqnarray}
  {\cal G}\,q_\alpha^c(x^0,\vec x)\,{\cal G}^{-1} &=&
  G_{\alpha\,\beta}^{-1}\,q_\beta^c(x'^0,\vec x\,'), \\
  {\cal G}\,Q_\alpha(y^0,\vec y)\,{\cal G}^{-1} &=&
  G_{\alpha\,\beta}^{-1}\,Q_\beta(y'^0,\vec y\,'),
\end{eqnarray}
where ${\cal G}$ is the Lorentz-boost operator, $G$ is its spinor representation,
and their definitions are given by
\begin{eqnarray}
  &&{\cal G}\,\left|X;\left(M_X,\vec 0\right)\{\ell\}\right>=
  \left|X;P_X\,\{\ell\}\right> \\
  &&G=\cosh{\frac{\varphi}{2}}+\alpha^3\sinh{\frac{\varphi}{2}},
  \qquad \beta=\tanh\varphi, \qquad 
  \gamma=\frac{1}{\sqrt{1-\beta^2}}=\cosh\varphi. \label{eq:Gdef}
\end{eqnarray}
We have assumed that the heavy meson is boosted in the $+z$ direction.
Here, we have $\vec \alpha=\gamma^0\vec \gamma$, and $\beta$ and $\gamma$ are the velocity 
and Lorentz factor of the heavy meson, respectively, which can be expressed 
in terms of one parameter, $\varphi$.

Even if two constituent quarks have equal times in one frame, they have different times in
another frame. In order to pull back the time of the heavy quark to that of the light
quark, we make an approximation that the heavy quark propagates freely over a short time interval,
assuming
\begin{equation}
  Q_\beta\left(t+\delta t,\vec y\right)\simeq
  \exp\left(-im_Q\gamma\delta t\right) Q_\beta\left(t,\vec y\right).
\end{equation}

Below we consider two cases. In the first case, the two constituent quarks have equal time
in the rest-frame of the meson, and we apply the above approximation to the heavy quark
in the moving frame. In the second case, the two constituents have equal times in the moving
frame of the meson, and we apply the approximation to the heavy quark in the rest-frame.
In these cases, different relations are derived between the wave function of the meson with
finite momentum and that of the meson at rest. These are as follows :
\begin{flushleft}
i) $t=x^0=y^0$ and $x'^0\ne y'^0$ (equal time in the rest-frame)
\end{flushleft}
\begin{eqnarray}
  &&x'^3=x^3\cosh\varphi+t\sinh\varphi,\;\quad
  x'^0=t\cosh\varphi+x^3\sinh\varphi, \nonumber \\
  &&y'^3=y^3\cosh\varphi+t\sinh\varphi,\;\quad 
  y'^0=t\cosh\varphi+y^3\sinh\varphi. \label{eq:cond_casei}
\end{eqnarray}
In this case, an approximate relation between the rest and moving frame wave
functions is given by
\begin{equation}
  {\psi_X^{\ell}}_{\alpha\,\beta}\left(\left(0,\vec x\right); P_X\right)\simeq
  G_{\alpha\,\gamma}G_{\beta\,\delta}\,{\psi_X^{\ell}}_{\gamma\,\delta}\left(
  \left(0,\vec x_\perp,\gamma^{-1}x^3\right); \left(M_X,\vec 0\right)\right)\,
  e^{i\left(M_X-m_Q\right)\gamma\,\beta\,x^3}. \label{eq:casei}
\end{equation}
\begin{flushleft}
ii) $t'=x'^0=y'^0$ and $x^0\ne y^0$ (equal time in the moving frame)
\end{flushleft}
\begin{eqnarray}
  &&x^3=x'^3\cosh\varphi-t'\sinh\varphi,\;\quad
  x^0=t'\cosh\varphi-x'^3\sinh\varphi, \nonumber \\
  &&y^3=y'^3\cosh\varphi-t'\sinh\varphi,\;\quad 
  y^0=t'\cosh\varphi-y'^3\sinh\varphi. \label{eq:cond_caseii}
\end{eqnarray}
In this case, an approximate relation between the rest and moving frame wave 
functions is given by
\begin{equation}
  {\psi_X^{\ell}}_{\alpha\,\beta}\left(\left(0,\vec x\right); P_X\right)\simeq
  G_{\alpha\,\gamma}G_{\beta\,\delta}\,{\psi_X^{\ell}}_{\gamma\,\delta}\left(
  \left(0,\vec x_\perp,\gamma x^3\right); \left(M_X,\vec 0\right)\right)\,
  e^{i\left(M_X-m_Q\right)\gamma\,\beta\,x^3}. \label{eq:caseii}
\end{equation}
The derivation of the Lorentz boosted wave functions given by Eqs.~(\ref{eq:casei})
and (\ref{eq:caseii}) is given in Appendix \ref{app:normal}.

%%%%%%%%%%%%%%%%%%%%%%%%%%%%%%%%
\subsection{Wave function}
\label{sub:wavefunction}
The explicit form of the wave function we obtained in the previous paper has the
form
\begin{subequations}
\label{eq:FWTpsi}
\begin{eqnarray}
  &&U_X(p)\psi _X^\ell\left(\vec x\right)=\psi_{X\; {\rm FWT}}^\ell
  (\vec x), \\
  &&U_X(p)\equiv U_c U_{\rm FWT}(p),\qquad \psi _X^\ell\left(\vec x\right)
  \equiv\psi _X^\ell\left(\left(0,\vec x\right);\left(M_X,\vec 0\right)\right),
  \\
  &&\psi_{X\; {\rm FWT}}^\ell(\vec x)=\psi_{X\,0}^\ell(\vec x)+
  \psi_{X\,1}^\ell(\vec x)+\ldots, \\
  &&\psi_{X\,0}^\ell(\vec x)=\Psi_\ell^+(\vec x)
%  \equiv\sqrt{M_X}\,\left( {\begin{array}{*{20}c}
%   0 & {\Psi _{j\,m}^k } \\ \end{array}} \right),
  \equiv\sqrt{M_X}\,\left( 
   0 ~~ {\Psi _{j\,m}^k } \right).
\end{eqnarray}
\end{subequations}
We have solved the Schr\"odinger equation in terms of the wave function
$\psi_{X\; {\rm FWT}}^\ell$, not $\psi_X^\ell$.
Here, the $4\times 4$ wave function, $\psi_X^\ell$, is transformed just once 
with the Foldy-Wouthuysen-Tani transformation, $U_{\rm FWT}$, acting on the
heavy quark and with the charge conjugation operator, $U_c$, into $\psi_{X\; 
{\rm FWT}}^\ell$, and its $i$-th order term in $1/m_Q$ is given by $\psi_
{X\,i}^\ell$. Higher-order corrections depend on the positive and negative
components of the heavy quark, $\Psi_\ell^\pm$, which are given by
\begin{equation}
  \Psi_\ell^+(\vec x) \equiv \sqrt{M_X}\,\left(
%  \begin{array}{*{20}c}
%  0 & \Psi _{j\,m}^k(\vec x) \\ \end{array} \right), \qquad
  0 ~~ \Psi _{j\,m}^k(\vec x)  \right), \qquad
  \Psi_\ell^-(\vec x) \equiv \sqrt{M_X}\,\left(
%  \begin{array}{*{20}c} \Psi _{j\,m}^k (\vec x)&0 \\ \end{array}\right).
  \Psi _{j\,m}^k (\vec x) ~~ 0 \right).
\end{equation}
Here, $\vec p=\vec p_Q$ appearing in the argument of $U_X(p)$ above is the
initial momentum of the heavy quark, and henceforth, the color index ($N_c=3$) is
omitted, because the form of the wave function is the same for all colors, $\ell$
stands for the set of quantum numbers, $j$, $m$, and $k$, and the
positive/negative component wave functions, $\Psi_\ell^\pm$, are given in terms
of a $4\times 2$ wave function, $\Psi _{j\,m}^k(\vec x)$, as
\begin{equation}
  \Psi _{j\,m}^k(\vec x) =\frac{1}{r}\left( {\begin{array}{*{20}c} u_k(r) \\
  -i\,v_k(r)\left(\vec n\cdot\vec \sigma\right) \\ \end{array}} \right)
  \;y_{j\,m}^k (\Omega),\label{eq:Psi+}
\end{equation}
where $r=\left|\vec x\right|$, $\vec n=\vec x/r$, $j$ is the total angular 
momentum of the meson, $m$ is its $z$ component, $k$ is a quantum number that 
takes only the values $k=\pm j,\;\pm(j+1)\;{\rm and}\;\ne 0$. (More details are
given in Appendix \ref{app:wf}.)

We showed in the previous paper that $\Psi_\ell^+(\vec x)$ is an eigenstate 
of the operator $K \equiv-\beta_q(\vec\Sigma_q\cdot\vec\ell+1)$, with
eigenvalue $k$, and we classified the spectra in terms of $k$. People 
normally classify the spectra in terms of $s_\ell^{~P}$, which is defined to be
an eigenvalue $s_\ell\,(s_\ell+1)$ of the operator $(\vec s_\ell)^2$,
\cite{IW91} together with parity, $P$.
We can show the following direct relation among $k$, $s_\ell$, and $P$;
i.e., $s_\ell$ and $P$ can be explicitly written only in terms of $k$ as
\cite{Rose61}
\begin{subequations}
\label{eq:sEllPi}
\begin{eqnarray}
  k &=& \pm \left( s_\ell + \frac{1}{2} \right) \quad {\rm or} \quad 
  s_\ell = \left| k \right| - \frac{1}{2}, \label{k} \\
  P &=& \frac{k}{|k|} (-1)^{|k| + 1},
\end{eqnarray}
\end{subequations}
where we have definitions
\begin{equation}
  \vec s_\ell\equiv \vec\ell+\frac{1}{2}\vec\Sigma_q,\quad
  (\vec s_\ell)^2=\vec\ell^{~2}+\vec\ell\cdot\Sigma_q+\frac{3}{4},
  \quad \vec\ell=-i\vec r \times \vec\nabla, \\
  \quad\vec\Sigma_q=\left(
  \begin{array}{*{20}c} \vec\sigma&0 \\ 0&\vec\sigma \\ \end{array} \right).
\end{equation}
The corresponding physical quantities between $k$ and $s_\ell$ 
and the first few states are listed in Tables \ref{table1} and \ref{table2},
respectively.

\begin{table}
\caption{Corresponding values for $k$, $s_\ell$, and $P$.}
\label{table1}
\begin{center}
\begin{tabular}
{@{\extracolsep{\fill}}|p{0.25in}|p{0.5in}|p{0.5in}|p{0.5in}|p{0.5in}|}
\hline \hline
  $P$  & $(-)^{j+1}$  & $(-)^j$ & $(-)^j$ & $(-)^{j+1}$ \\ \hline
%%%%%%%%%%%%%%%%%%%%%%%%%%%
  $k$  & $-(j+1)$ & $j+1$ & $-j$ & $j$ \\ \hline
%%%%%%%%%%%%%%%%%%%%%%%%%%%
  $s_\ell$ & $j+\frac{1}{2}$ & $j+\frac{1}{2}$ & $j-\frac{1}{2}$ & 
  $j-\frac{1}{2}$ \\ \hline  \hline
%%%%%%%%%%%%%%%%%%%%%%%%%%%
  $\Psi_j^k$ & $\Psi_j^{-(j+1)}$ & $\Psi_j^{j+1}$ & $\Psi_j^{-j}$ & $\Psi_j^j$
  \\
\hline
\end{tabular}
\end{center}
\end{table}
\begin{table}
\caption{States classified with respect to various quantum numbers.}
\label{table2}
\begin{center}
\begin{tabular}
{@{\extracolsep{\fill}}|p{0.25in}|p{0.25in}|p{0.25in}|p{0.25in}|
p{0.25in}|p{0.25in}|p{0.25in}|p{0.25in}|p{0.25in}|p{0.25in}|p{0.25in}|}
\hline \hline
  $j^P$         & $0^-$    & $1^-$     & $0^+$     & $1^+$    &
   $1^+$      & $2^+$     & $1^-$     & $2^-$ &  $2^-$ & $2^+$  \\ \hline
%%%%%%%%%%%%%%%%%%%%%%%%%%%
  $k$           &    -1    &    -1     &     1     &     1    &
      -2      &    -2     &     2     &     2    & -3 & 3\\ \hline
%%%%%%%%%%%%%%%%%%%%%%%%%%%
  $s_\ell^P$ & $\frac{1}{2}^-$ & $\frac{1}{2}^-$ & $\frac{1}{2}^+$ & 
  $\frac{1}{2}^+$ & $\frac{3}{2}^+$ & $\frac{3}{2}^+$ & $\frac{3}{2}^-$ &
  $\frac{3}{2}^-$
  & $\frac{5}{2}^-$ & $\frac{5}{2}^+$\\ \hline  \hline
%%%%%%%%%%%%%%%%%%%%%%%%%%%
  $\Psi_j^k$ & $\Psi_0^{-1}$ & $\Psi_1^{-1}$ & $\Psi_0^1$ & $\Psi_1^1$ &
  $\Psi_1^{-2}$ & $\Psi_2^{-2}$ & $\Psi_1^2$ & $\Psi_2^2$ &
  $\Psi_2^{-3}$ & $\Psi_2^3$ \\
\hline
\end{tabular}
\end{center}
\end{table}

From Tables \ref{table1} and \ref{table2}, it is seen that there are both advantages and
disadvantages of using these quantities to classify the heavy mesons. The quantity
$s_\ell^P$ has an intuitive meaning that it represents the
total angular momentum of the light degrees of freedom, while $k$ does not. 
However, $k$ includes information concerning parity, while $s_\ell$ does not. Hence
we need to add $P$, writing it as $s_\ell^P$. 
Certainly either one can be used to classify the heavy meson 
spectra equally well; i.e., they are equivalent.
\subsection{Normalization}
\label{sub:normal}
Normalization of the state $\left|X;P_X\,\{\ell\}\right>$ is formally given
by
\begin{equation}
  \left<X;P_X'\,\{\ell'\}\right.\left|X;P_X\,\{\ell\}\right>
  =(2\pi)^3\,2{P_X}_0\,\delta^3\left(\vec P_X-\vec P_X'\right)\delta_
  {\ell',\ell} \label{eq:norm1}
\end{equation}
or
\begin{equation}
  \int d^3z\, {\rm tr}\left[{\psi_{X}^{\ell'}}^\dagger\left(\left(0,
  \vec z\right);P_{X}\right)\psi_X^\ell\left(\left(0,\vec z\right);P_X\right)
  \right]\simeq 2{P_X}_0\,\delta_{\ell\,\ell'}, \label{eq:norm2}
\end{equation}
whose derivation is given in Appendix \ref{app:normal}. Actually, the 
normalization given by Eq.~(\ref{eq:norm2}) does not hold, because of 
approximations adopted in the previous subsection, and hence, instead of
Eq.~(\ref{eq:norm2}), we define the normalization in the rest-frame as 
follows :
\begin{equation}
  \int\, d^3z\,{\rm tr}\left[{\psi_{X}^{\ell'}}^\dagger\left(\left(0,\vec z
  \right);\left(M_X,\vec 0\right)\right)\psi_X^\ell\left(\left(0,\vec z
  \right);\left(M_X,\vec 0\right)\right)\right]= 
  2M_X\,\delta_{\ell\,\ell'}. \label{eq:norm3}
\end{equation}
In order to calculate the normalization in a moving frame, we have to specify
the quantum numbers in which we are interested. Because we would like to compute 
the Isgur-Wise function and its higher orders, the cases of the pseudoscalar 
state $0^-$ and vector state $1^-$ are calculated in the following. 

The normalization condition given by Eq.~(\ref{eq:norm3}) can be rewritten 
in terms of $\psi_{X\; {\rm FWT}}^\ell(\vec x)$ as
\begin{equation}
  \int d^3x\,{\rm tr}\left[{\psi_{X\,{\rm FWT}}^{\ell'}}^\dagger\left(\vec x
  \right)\psi_{X\,{\rm FWT}}^\ell\left(\vec x\right)\right]= 
  2M_X\,\delta_{\ell\,\ell'}. \label{eq:norm4}
\end{equation}
All the details concerning the wave functions are given in Appendix \ref{app:wf}.

In order to calculate the normalization of the wave function in a moving
frame and/or the matrix elements of the semi-leptonic decay processes,
we need to develop a formulation of their calculations in terms of the rest-frame
wave functions. Let us assume that a physical quantity is already given
in terms of the rest-frame wave functions and rewrite them in terms of the
FWT-transformed ones as
\begin{eqnarray}
  &&\int d^3x\, \left({\cal O}_q\right)_{\alpha\,\alpha'}
  \left({\cal O}_Q\right)_{\beta\,\beta'}A_{\alpha\,\gamma}^{'\,*} 
  B_{\beta\,\delta}^{'\,*} {\psi_{X'}^{\ell\,'\,*}}_{\gamma\,\delta}
  A_{\alpha'\,\gamma'}B_{\beta\,\delta'}{\psi_X^\ell}_{\gamma'\,\delta'}
  \nonumber \\
  &&=\int d^3x\, {\rm tr}\left(\psi_{X'}^{\ell\,'\,\dagger} A^{'\,\dagger} 
  {\cal O}_q A \psi_X^\ell B^{\rm T}{\cal O}_Q^{\rm T}B^{'\,*}\right) 
  \nonumber \\
  &&=\int d^3x\, {\rm tr}\left[\psi_{X'\,{\rm FWT}}^{\ell\,'\,\dagger} 
  A^{'\,\dagger} {\cal O}_q A\psi_{X\,{\rm FWT}}^\ell U_X^{-1}(p_Q) B^{\rm T}
  {\cal O}_Q^{\rm T}B^{'\,*}\,U_{X'}(p'_Q) \right], \label{eq:matrixFWT}
\end{eqnarray}
where $({\cal O}_q)_{\alpha\,\alpha'}$, $A'_{\alpha\,\gamma}$, and 
$A_{\alpha\,\gamma'}$ act on light quarks, while 
$({\cal O}_Q)_{\beta\,\beta'}$, $B'_{\beta\,\delta}$ and $B_{\beta\,\delta'}$
act on heavy quarks, and use has been made of
\begin{eqnarray}
  &&\psi_X^\ell=U_X^{-1}(p_Q)\otimes \psi_{X\,{\rm FWT}}^\ell(p_Q)
  =\psi_{X\,{\rm FWT}}^\ell\,U_X^{-1}(p_Q), \quad
  \psi_{X'}^\ell=\psi_{X\,'\,{\rm FWT}}^\ell\,U_X^{-1}(p'_Q), \nonumber \\
  &&U_X^{-1}(p_Q) B^{\rm T}{\cal O}_Q^{\rm T}B^{'\,*}\, U_{X'}(p'_Q)
  =U_{X\,{\rm FWT}}^{-1}(p_Q) U_c^{-1} B^{\rm T}{\cal O}_Q^{\rm T}B^{'\,*}\, 
  U_c U_{X\,'\,{\rm FWT}}(p'_Q).
\end{eqnarray}

In the course of obtaining relativistic results from the rest-frame wave 
functions, we calculate the normalization of the wave function in a moving 
frame in two extreme cases, i)~$t=x^0=y^0$ and ii)~$t'=x'^0=y'^0$.
The FWT-transformed rest-frame wave function is given by, to first 
order in $1/m_Q$, \cite{Matsuki97}
\begin{subequations}
\label{eq:psi01}
\begin{eqnarray}
  \psi_{X\,{\rm FWT}}^\ell(0^-)&=&\Psi_{-1}^++c_{1-}^{-1,\,1}\Psi_1^-
  +O\left(1/m_Q^2\right), \label{eq:psi0-}\\
  \psi_{X\,{\rm FWT}}^\ell(1^-)&=&\Psi_{-1}^++c_{1+}^{-1,\,2}\Psi_2^+
  +c_{1-}^{-1,\,1}\Psi_1^-+c_{1-}^{-1,\,-2}\Psi_{-2}^-+O\left(1/m_Q^2\right),
  \label{eq:psi1-}
\end{eqnarray}
\end{subequations}
where only the value of the $k$ quantum number is written as a subscript of 
$\Psi_\ell^\pm$. The total angular momentum, $j$, though not given explicitly,
should be the same on both sides of the equations and the coefficients, 
$c_{1\pm}^{k,\,k'}$, are given in \citen{Matsuki97}. Their explicit expressions 
are not necessary here, since we show that the higher-order
corrections do not contribute to the physical quantities at this order, 
$O(1/m_Q)$. Actually, Eq.~(\ref{eq:psi01}) can be derived 
from the conservation of total angular momentum and parity alone, without
explicit interaction terms specified, since we know the complete set of
eigenfunctions with $j$, $m$, and $k$ quantum numbers. (Details are given in
Appendix \ref{app:wf}.)

We calculate the normalizatio up to first-order in $1/m_Q$ below.
\begin{flushleft}
i) $t=x^0=y^0$ and $x'^0\ne y'^0$ (equal time in the rest-frame)
\end{flushleft}
Using Eqs.~(\ref{eq:casei}), (\ref{eq:norm4}) and (\ref{eq:matrixFWT}), we obtain,
to first-order in $1/m_Q$ and both for $0^-$ and $1^-$ states given 
by Eq.~(\ref{eq:psi01}),
\begin{equation}
  \int\; d^3x\; {\rm tr}\left[\psi^{\ell'\,\dagger}_X\left(\left(0,\vec x
  \right);P_X\right)\psi^\ell_X\left(\left(0,\vec x\right);P_X\right)
  \right]=2M_X\gamma^3+O\left(1/m_Q^2\right). \label{eq:normP1}
\end{equation}
\begin{flushleft}
ii) $t'=x'^0=y'^0$ and $x^0\ne y^0$ (equal time in the moving frame)
\end{flushleft}
Similarly using Eq.~(\ref{eq:caseii}), (\ref{eq:norm4}) and (\ref{eq:matrixFWT}),
we obtain
\begin{equation}
  \int d^3x\; {\rm tr}\left[\psi^{\ell'\,\dagger}_X\left(\left(0,\vec x
  \right);P_X\right)\psi^\ell_X\left(\left(0,\vec x\right);P_X\right) 
  \right]=2M_X\gamma+O\left(1/m_Q^2\right). \label{eq:normP2}
\end{equation}
Only this case agrees with the relativistic normalization given by 
Eq.~(\ref{eq:norm2}).

\subsection{Matrix elements}
\label{sub:elements}
The light anti-quark is treated as a spectator; that is, it does not 
interact with other particles, except for gluons represented by potentials
in our model. The heavy quark has a weak vertex, and its current is 
in general given by
\begin{equation}
  j(t,\vec x)={Q_{X'}^\dagger}_\alpha(t,\vec x)\,{\cal O}_{\alpha\,\beta}
  \,{Q_X}_\beta(t,\vec x),\label{eq:current}
\end{equation}
where ${\cal O}$ is a $4\times 4$ matrix. By inserting the number operator
of the anti-quark and by assuming vacuum dominance among intermediate
states, we obtain the formula for the matrix element of the above current,
Eq.~(\ref{eq:current}), between heavy mesons :
\begin{eqnarray}
  &&\left<X';P_{X'}\,\{\ell'\}\left|j(t,\vec x)
  \right|X;P_X\,\{\ell\}\right>
  \nonumber \\
  &&\simeq \int\,d^3y\,{\rm tr}\left[
  {\psi_{X'}^{\ell'}}^\dagger\left(\left(0,\vec y-\vec x
  \right);P_{X'}\right)\left({\cal O}\otimes\psi_X^\ell\left(\left(0,\vec y-
  \vec x\right);P_X\right)\right)\right]
  \nonumber \\
  &&\quad \times\, e^{-i\left(P_X-P_{X'}\right)\cdot x} \label{eq:mat1}
\end{eqnarray}
or
\begin{eqnarray}
  &&\left<X';P_{X'}\,\{\ell'\}\left|j(0,\vec 0)
  \right|X;P_X\,\{\ell\}\right>
  \nonumber \\
  &&\simeq \int\,d^3x\,{\rm tr}\left[{\psi_{X'}^{\ell'}}^\dagger\left(
  \left(0,\vec x\right);P_{X'}\right)\,\psi_X^\ell\left(\left(0,\vec x
  \right);P_X\right)\,{\cal O}^{\rm T}\right].\label{eq:mat2}
\end{eqnarray}

Now we would like to evaluate the matrix element of the process $\bar B
\rightarrow D^{(*)}\ell\bar\nu_\ell$ in two frames, the
$\bar B$ rest-frame and the Breit frame. (The velocities of $\bar B$ and $D$
have the same magnitude and have the opposite directions.)  In each frame, we
apply two approximate relations, Eqs.~(\ref{eq:cond_casei}) and
(\ref{eq:cond_caseii}). As mentioned in the previous subsection, the
normalization condition Eq.~(\ref{eq:norm2}) of the wave function holds in
the approximation ii) (equal time in the moving frame) but not in the
approximation i) (equal time in the rest-frame).  In the approximation i), we
need to renormalize the matrix elements by normalizing the wave
function, and we redefine the matrix elements of the current as follows :
\begin{eqnarray}
  &&\left<D^{(*)}\left|j(0,{\vec 0})\right|\bar B\right>
  \nonumber \\
  &&=\frac{
  {\sqrt{2P_{{\bar B}_0} 2P_{{D^{(*)}}_0}}\, \int d^3x\;{\rm tr}
  \left[\psi^{\ell'\,\dagger}_{D^{(*)}}\left(\left(0,\vec x\right);P_{D^{(*)}
  }\right)\,\psi^{\ell}_{\bar B}(\left (0,\vec x\right);P_{\bar B}) 
  {\cal O}^{\rm T}\right] }}% \over  
  {{\sqrt{\int d^3x\;
  {\rm tr}\left|\psi^{\ell'}_{D^{(*)}}(\left(0,\vec x\right);P_{D^{(*)}}
  )\right|^2\,\int d^3x\;{\rm tr} \left|\psi^{\ell}_{\bar B}(\left(0,\vec x
  \right);P_{\bar B})\right|^2}}}.
\end{eqnarray}
The numerator here is calculated below using the results of
\S \ref{sub:lorentz}, while the denominator is calculated in
\S \ref{sub:normal} and given by
Eqs.~(\ref{eq:normP1}) and (\ref{eq:normP2}).

\begin{flushleft}
1) $\bar B$ rest-frame \\
1-i) $t=x^0=y^0$ for the $D^{(*)}$ meson
\end{flushleft}
In this case, the matrix element is given by, with our approximations,
\begin{eqnarray}
  &&\int d^3x\; {\rm tr}\left[\psi^{\ell'\,\dagger}_{D^{(*)}}
  \left(\left(0,\vec x\right);P_{D^{(*)}}\right)\,\psi^{\ell}_{\bar B}(\left
  (0,\vec x\right);P_{\bar B}) {\cal O}^{\rm T}\right]
  \nonumber \\
  &&=\int\,d^3x\,G_{\gamma\,\delta}^*G_{\alpha\,
  \epsilon}^*\,{\psi_{D^{(*)}}^{\ell'}}_{\delta\,\epsilon}^*\left(\left(0,
  \vec x_\perp,\gamma^{-1} x^3\right);\left(M_{D^{(*)}},\vec 0\right)\right) 
  {\cal O}_{\alpha\,\beta}\,{\psi_{\bar B}^\ell}_{\gamma\,\beta}
  \left(\vec x\right)
  \nonumber \\
  &&\quad\times\, e^{-i\left(M_{D^{(*)}}-m_c\right)\gamma\,\beta\, x^3}.
  \label{eq:me1}
\end{eqnarray}
\begin{flushleft}
1-ii) $t'=x'^0=y'^0$ for the $D^{(*)}$ meson
\end{flushleft}
In this case, the matrix element is given by
\begin{eqnarray}
  &&\int d^3x\; {\rm tr}\left[\psi^{\ell'\,\dagger}_{D^{(*)}}
  \left(\left(0,\vec x\right);P_{D^{(*)}}\right)\,\psi^{\ell}_{\bar B}(\left
  (0,\vec x\right);P_{\bar B}) {\cal O}^{\rm T}\right]
  \nonumber \\
  &&=\int\,d^3x\,G_{\gamma\,\delta}^*G_{\alpha\,
  \epsilon}^*\,{\psi_{D^{(*)}}^{\ell'}}_{\delta\,\epsilon}^*\left(\left(0,
  \vec x_\perp,\gamma x^3\right);\left(M_{D^{(*)}},\vec 0\right)\right)
  {\cal O}_{\alpha\,\beta}\,{\psi_{\bar B}^\ell}_{\gamma\,\beta}
  \left(\vec x\right)
  \nonumber \\
  &&\quad\times \, e^{-i\left(M_{D^{(*)}}-m_c\right)\gamma\,\beta\, x^3}.
  \label{eq:me2}
\end{eqnarray}
\begin{flushleft}
2) Breit frame \\
2-i) $t=x^0=y^0$ for the $\bar B$ and $D^{(*)}$ mesons
\end{flushleft}
In this case, the matrix element is given by, with our approximations,
\begin{eqnarray}
  &&\int d^3x\; {\rm tr}\left[\psi^{\ell'\,\dagger}_{D^{(*)}}
  \left(\left(0,\vec x\right);P_{D^{(*)}}\right)\,\psi^{\ell}_{\bar B}(\left
  (0,\vec x\right);P_{\bar B}) {\cal O}^{\rm T}\right]
  \nonumber \\
  &&=\gamma\,\int\,d^3x\,{\psi_{D^{(*)}}^{\ell'}}_
  {\gamma\,\alpha}^*\left(\vec x\right) 
  \left(G^\dagger{\cal O}G^{-1}\right)_{\alpha\,\beta}
  {\psi_{\bar B}^\ell}_{\gamma\,\beta}\left(\vec x\right) 
  \nonumber \\
  &&\times\, 
  \exp\left[{-\frac{i}{2}\left(M_{D^{(*)}}-m_c+M_B-m_b \right)\gamma^2\,\beta\,x^3}\right].
  \label{eq:me3}
\end{eqnarray}
\begin{flushleft}
2-ii) $t'=x'^0=y'^0$ for the ${\bar B}$ and $D^{(*)}$ mesons
\end{flushleft}
In this case, the matrix element is given by
\begin{eqnarray}
  &&\int d^3x\; {\rm tr}\left[\psi^{\ell'\,\dagger}_{D^{(*)}}
  \left(\left(0,\vec x\right);P_{D^{(*)}}\right)\,\psi^{\ell}_{\bar B}(\left
  (0,\vec x\right);P_{\bar B}) {\cal O}^{\rm T}\right]
  \nonumber \\
  &&=\gamma^{-1}\,\int\,d^3x\,{\psi_{D^{(*)}}^
  {\ell'}}_{\gamma\,\alpha}^* \left(\vec x\right)
  \left(G^\dagger{\cal O}G^{-1}\right)_{\alpha\,\beta}
  {\psi_{\bar B}^\ell}_{\gamma\,\beta}\left(\vec x\right) 
  \nonumber \\
  &&\times \,
  \exp\left[{-\frac{i}{2}\left(M_{D^{(*)}}-m_c+M_B-m_b \right)\,\beta\,x^3}\right].
  \label{eq:me4}
\end{eqnarray}
\section{Semi-Leptonic weak form factor}
\label{sec:IW}
Using the results derived in the previous section, we can now calculate 
semi-leptonic weak form factors including the Isgur-Wise function. In this 
case the currents are given by
\begin{subequations}
\label{eq:currents}
\begin{eqnarray}
  j_\mu={c^\dagger}(t,\vec x)\,\gamma^0\gamma_\mu\,b\,(t,\vec x), \\
  j_\mu^5={c^\dagger}(t,\vec x)\,\gamma^0\gamma_\mu\gamma_5\,b\,(t,\vec x).
\end{eqnarray}
\end{subequations}
Define the six independent semi-leptonic weak form factors as \cite{POL}
\begin{subequations}
\label{eq:6form}
\begin{eqnarray}
  \left<D\right|j_\mu(0,{\vec 0})\left|\bar B\right>
  &=&\sqrt{M_B M_D}\left(\xi_+(\omega)(v_{\bar B}+v_D)_\mu+\xi_-(\omega)
  (v_{\bar B}-v_D)_\mu\right), \label{eq:wff1} \\
  \left<D^*\right|j_\mu(0,{\vec 0})\left|\bar B\right>
  &=&i\;\sqrt{M_B M_{D^*}}\xi_V(\omega)\;\epsilon_{\mu\nu\rho\sigma}
  \epsilon^{*\,\nu}v_{D^*}^\rho\;v_{\bar B}^\sigma, \label{eq:wff2} \\
  \left<D^*\right|j_\mu^5(0,{\vec 0})\left|\bar B\right>
  &=&\sqrt{M_B M_{D^*}}\left((1+\omega)\xi_{A_1}(\omega)\;\epsilon_\mu^* 
  \right.\nonumber \\
  \quad &&\left.-\xi_{A_2}(\omega)\;\left(\epsilon^*\cdot v_{\bar B}\right) 
  v_{\bar B\;\mu}-\xi_{A_3}(\omega)\;\left(\epsilon^*\cdot v_{\bar B}\right) 
  v_{D^*\;\mu}\right),\label{eq:wff3}
\end{eqnarray}
\end{subequations}
with 
\begin{equation}
  v_X^\mu={P_X^\mu/ M_X},\quad{\rm for}~X={\bar B},~D^{(*)},\qquad 
  \omega=v_{\bar B}\cdot v_D.
\end{equation}
\subsection{Zeroth order (Isgur-Wise function)}
\label{sub:0th}
After some calculations, all the above form factors are found to be proportional to the
Isgur-Wise function, $\xi(\omega)$, or vanish and are given by
\begin{subequations}
\label{eq:6xi}
\begin{eqnarray}
  &&\xi_+(\omega)=\xi_V(\omega)=\xi_{A_1}(\omega)=\xi_{A_3}(\omega)
    =\xi(\omega),\\
  &&\xi_-(\omega)=\xi_{A_2}(\omega)=0. \label{eq:xiall}
\end{eqnarray}
\end{subequations}
Below we only briefly show how to calculate $\xi_+(\omega)=\xi(\omega)$ 
from Eq.~(\ref{eq:wff1}), since the other functions are obtained similarly. All the 
details are given in Appendix \ref{app:IW}. We define the Isgur-Wise 
function as
\begin{eqnarray}
  \left<D\left|j_0(0,{\vec 0})\right|\bar B\right>&=&\frac{
  {\sqrt{2P_{\bar B\,0} 
  2P_{D\,0}}\int d^3x\;{\rm tr}\left[\psi^{\ell'\,\dagger}_{D\,0}\left(\left(0,
  \vec x\right);P_D\right)\,\psi^{\ell}_{\bar B\,0}\left(\left(0,\vec x\right)
  ;P_{\bar B}\right)\right] }}%\over
  {{\sqrt{\int d^3x\; {\rm tr}\left|\psi^{\ell'}_{D\,0}(\left(0,
  \vec x\right);P_D)\right|^2\,\int d^3x\; {\rm tr}\left|\psi^{\ell}_
  {\bar B\,0}(\left(0,\vec x\right);P_{\bar B})\right|^2}}}
  \nonumber \\
  &=&\sqrt{M_B M_D}\xi(\omega)(v_{\bar B}+v_D)_0. \label{eq:isgur1}
\end{eqnarray}
The expression Eq.~(\ref{eq:isgur1}) can be further reduced in two cases of the
$\bar B$ frame as follows.
\begin{flushleft}
1) $\bar B$ rest-frame
\end{flushleft}
In this frame, $\bar B$ is at rest, $D$ is moving in the $+z$ direction with
velocity $\beta$ and $\xi(\omega)$ is given by 
\begin{equation}
  \xi(\omega)=\frac{ {\sqrt{2\omega}\,\int d^3x\; {\rm tr}\left[\psi^{\ell'\,
  \dagger}_{D\,0}\left(\left(0,\vec x\right);P_D\right)\psi^{\ell}
  _{\bar B\,0}\left(\vec x\right)\right] }} %\over
  {\sqrt{M_B}(1+\omega)
  \sqrt{\int d^3x\; {\rm tr}\left|\psi^{\ell'\,\dagger}_{D\,0}
  \left(\left(0,\vec x\right);P_D\right)\right|^2}},
\end{equation}
with the relations
\begin{eqnarray}
  &&P_{\bar B\,0}=M_B, \qquad P_{D\,0}=M_D v_{D\,0}=M_D\omega, \qquad 
  v_{\bar B\,0}=1, \nonumber \\
  &&v_{D\,0}=v_{\bar B}\cdot v_D=\gamma=\frac{1}{\sqrt{1-\beta^2}}=\omega,
  \qquad \beta=\frac{\sqrt{\omega^2-1}}{\omega}.\label{eq:rel1}
\end{eqnarray}
\begin{flushleft}
2) Breit frame
\end{flushleft}
In this frame, $\bar B$ is moving in the $+z$ direction, $D$ in
the $-z$ with the same velocity $\beta/2$, and $\xi(\omega)$ is given by
\begin{equation}
  \xi(\omega) = {\int d^3x\;{\rm tr}\left[\psi^{\ell'\,\dagger}_{D\,0}
  \left(\left(0,\vec x\right);P_D\right)
  \psi^\ell_{{\bar B}\,0}\left(\left(0,\vec x
  \right);P_{\bar B}\right)\right]\over \sqrt{\int d^3x\; {\rm tr}\left|\psi^
  {\ell'}_{D\,0}\left(\left(0,\vec x\right);P_D\right)\right|^2\,
  \int d^3x\; {\rm tr}\left|\psi^\ell_{{\bar B}\,0}
  \left(\left(0,\vec x\right);P_{\bar B}
  \right)\right|^2}},
\end{equation}
with the relations
\begin{eqnarray}
  &&P_{\bar B\,0}=M_{\bar B} \gamma, \qquad P_{D\,0}=M_D \gamma, \qquad 
  v_{\bar B\,0}=v_{D\,0}=\gamma={1\over \sqrt{1-(\beta/2)^2}},\nonumber\\
  &&v_{\bar B}\cdot v_D=\gamma^2\left(1+\frac{\beta^2}{4}\right)=\omega,
  \qquad \gamma=\sqrt{\omega+1\over2},\qquad {\beta\over2}=\sqrt{\omega-1
  \over \omega+1}. \label{eq:rel2}
\end{eqnarray}
We note that here and below we have used the same notation 
for the velocity and Lorentz factor both in the $\bar B$ rest-frame and the Breit frame,
although they are different quantities, as one can see from Eqs.~(\ref{eq:rel1}) and 
(\ref{eq:rel2}).

We calculate this Isgur-Wise function in four cases below. The details are
given in Appendix \ref{app:sub:first}.
\begin{flushleft}
1-i) $\bar B$ rest-frame and $t=x^0=y^0$. In this case, we have
\end{flushleft}
\begin{equation}
  \xi(\omega)={1\over\omega}-{1\over6}\beta^2\omega{\tilde E_D}^2
  \left<r^2\right>+{\beta^2\over 4}+ \; O(\beta^4),
\end{equation}
where
\begin{equation}
  \left<r^2\right>\equiv \int d^3x\,{\rm tr}\left[{\psi_{X\,0}^\ell}^\dagger
  (\vec x)r^2\psi_{X\,0}^\ell(\vec x)\right], 
  \qquad {\rm for}~X=D,~{\rm or}~{\bar B},
\end{equation}
both of which give the same result, since this is the lowest-order 
calculation.
\begin{flushleft}
1-ii) $\bar B$ rest-frame and $t'=x'^0=y'^0$ :
\end{flushleft}
\begin{equation}
  \xi(\omega)=1-{1\over6}\beta^2\omega{\tilde E_D}^2\left<r^2\right>-
  {\beta^2\over4}+\;O(\beta^4).
\end{equation}
\begin{flushleft}
2-i) Breit frame and $t=x^0=y^0$ :
\end{flushleft}
\begin{equation}
  \xi(\omega)=\gamma^{-2}-{1\over6}\left({\beta\over2}\right)^2\gamma^2({\tilde E_{\bar B}}
  +{\tilde E_D})^2 \left<r^2\right>+O(\beta^4).
\end{equation}
\begin{flushleft}
2-ii) Breit frame and $t'=x'^0=y'^0$ :
\end{flushleft}
\begin{equation}
  \xi(\omega)=\gamma^{-2}-{1\over6}\left({\beta\over2}\right)^2\gamma^{-2}
  ({\tilde E_{\bar B}} + {\tilde E_D})^2\left<r^2\right>+O(\beta^4).
\end{equation}
In the above we have used
\begin{equation}
  \tilde E_D\equiv M_D-m_c, \qquad \tilde E_{\bar B}\equiv M_B-m_b. \label{diffXx}
\end{equation}
The rhs of Eq.~(\ref{diffXx}) can be expanded in $1/m_Q$ as
\begin{equation}
  \tilde E_X = M_X - m_Q=\left(m_Q+\sum_{i=0} \frac{C_X^i}{(m_Q)^i} \right) -m_Q
  = \sum_{i=0} \frac{C_X^i}{(m_Q)^i}, \label{C_X}
\end{equation}
where the {\it i}-th order in $1/m_Q$ of $M_X$ is given by $C_X^i/(m_Q)^i$, 
and it is shown in Ref. \citen{Matsuki97} that $C_X^i$ depends only on the light 
quark mass $m_q$, and hence $C_D^i=C_B^i=C^i(m_q)$ in our case, since light quarks
are either $u$ or $d$, and we have set $m_u=m_d$. Equation (\ref{C_X}) is for the $0^-$
state. When it is written for the $1^-$ state, the coefficients are given by ${C_X^i}'$.
All the expressions obtained above have the same form, up to first-order
in $\omega$ in the vicinity of $\omega=1$, and up to the zeroth-order in
$1/m_Q$, as given by
\begin{eqnarray}
  && \xi(\omega)=1-\left({1\over 2}+{1\over 3}{\bar \Lambda}^2\left<r^2\right>
  \right)\left(\omega-1\right),  \label{eq:xi} \\
  && {\bar \Lambda}=\lim_{m_Q\rightarrow \infty}\tilde E_D=
  \lim_{m_Q\rightarrow \infty}\tilde E_{\bar B}=C^0, \label{eq:Lambda}
\end{eqnarray}
that is,
\begin{equation}
  \xi(1)=1,\qquad \xi'(1)=-{1\over 2}-{1\over 3}{\bar \Lambda}^2
  \left<r^2\right>.
\end{equation}
The form for $\xi(\omega)$ given by Eq.~(\ref{eq:xi}) is derived from our 
semi-relativistic formulation and coincides with those derived from other 
considerations, \cite{HS,AMT,VO} but we believe that our derivation is the
most general and does not depend on any specific model. This form
gives the lower bound $-1/2$ for $\xi'(1)$ or the upper bound $1/2$ for 
$-\xi'(1)$, as shown in Refs. \citen{HS}, \citen{AMT} and \citen{VO}. All the 
other form factors, $\xi_-(\omega)$, $\xi_V(\omega)$, and $\xi_{A_i}(\omega)$ 
for $i=1 - 3$, are similarly calculated and given by Eq.~(\ref{eq:6xi})
up to this order, $(1/m_Q)^0$.

Using the values for the parameters obtained in Ref. \citen{Matsuki97} 
to calculate the mass spectra of heavy mesons, we can calculate $\xi'(1)$ to
first and second orders in $1/m_Q$ as
\begin{equation}
  \xi'(1)=-1.44,
\end{equation}
%
% \bar\Lambda=0.412 GeV, <r^2>=5.663 GeV^{-2} in 1998
% \bar\Lambda=0.7522 GeV, <r^2>=5.009 GeV^{-2} in 2007
% xi'(1)=-1.4447
%
where we have used the values $\bar\Lambda=0.752~{\rm GeV}$ and
$\left<r^2\right>=5.009~{\rm GeV}^{-2}$, taken from Ref. \citen{Matsuki05}.
The values of $\xi'(1)$ given in references are listed in Ref. \citen{AMT}.
\subsection{First order}
\label{sub:1st}
There is only one kind of the first-order correction, i.e., that from the 
wave functions. The first-order corrections to the form factors
defined by Eq.~(\ref{eq:6form}) from the wave functions are 
straightforwardly calculated, although the calculation is cumbersome. 
Hence, here we do not repeat calculations similar to \S
\ref{sub:0th} and give only the final results below. 

We have calculated all four cases,
1-i) - 2-ii). The results show that there are no contributions from the first-order
corrections of the wave functions to the form factors. Although we attempted 
to obtain relativistically invariant results from knowledge of the rest frame by 
applying four different Lorentz-boost frames, the first-order corrections 
in all cases are not invariant, except one, the case 2-ii). The Breit frame 
with $t'=x'^0=y'^0$ gives relativistic results that
coincide with those of Ref. \citen{POL}. Hence, the following results could be
a conjecture from our model, but we believe that our results are very plausible,
since they agree with relativistic results
which come from a semi-relativistic potential model. Brief 
derivations of these form factors are given in Appendix \ref{app:sub:second}.

As stressed in the Introduction, our approach does not use fields, and hence
there is no other kind of first order, i.e., the currents cannot be
expanded in $1/m_Q$ in terms of the effective velocity-dependent fields,
as in the HQET given in many papers as
\[
  j_\mu=c_v^\dagger\gamma^0\gamma_\mu b_v+c_v^\dagger\left(\frac{-i}{2m_c}\gamma^0
  \overleftarrow{D{\kern -8pt /}}\gamma_\mu+\frac{i}{2m_b}\gamma^0\gamma_\mu
  \overrightarrow{D{\kern -8pt /}}\right)\,b_v.
\]

Six form factors including zeroth-order terms for completeness with the decomposition
of Neubert and Rieckert \cite{POL} are given by
\begin{subequations}
\label{eq:6form1}
\begin{eqnarray}
  \xi_+(\omega)&=&\left[1+\left({1\over m_c}+{1\over m_b}\right)\rho_1(\omega)
    \right]\xi(\omega), \label{eq:6form1a} \\
  \xi_-(\omega)&=&\left[-{\bar \Lambda\over 2}+\rho_4(\omega)\right]
    \left({1\over m_c}-{1\over m_b}\right)\xi(\omega), \\
  \xi_V(\omega)&=&\left[1+{\bar \Lambda\over 2}\left({1\over m_c}+{1\over m_b}
    \right)+{1\over m_c}\rho_2(\omega)+{1\over m_b}\left(\rho_1(\omega)
    -\rho_4(\omega)\right)\right]\xi(\omega), \\
  \xi_{A_1}(\omega)&=&\left[1+{\bar \Lambda\over 2}{\omega-1\over \omega+1}
    \left({1\over m_c}+{1\over m_b}\right)+{1\over m_c}\rho_2(\omega)\right. \nonumber \\
  \quad &&\left.+ {1\over m_b}\left(\rho_1(\omega)-{\omega-1\over \omega+1}\rho_4(\omega)
    \right)\right]\xi(\omega), \\
  \xi_{A_2}(\omega)&=&{1\over \omega+1}{1\over m_c}\left[-{\bar \Lambda}
    +(\omega+1)\rho_3(\omega)-\rho_4(\omega)\right]\xi(\omega), \\
  \xi_{A_3}(\omega)&=&\left[1+{\bar \Lambda\over 2}\left({\omega-1\over 
    \omega+1}{1\over m_c}+{1\over m_b}\right)+{1\over m_c}\left(\rho_2(\omega)
    -\rho_3(\omega)-{1\over \omega+1}\rho_4(\omega)\right)\right. \nonumber \\
  \quad &&\left.+{1\over m_b}\left(\rho_1(\omega)-\rho_4(\omega)\right)\right]
  \xi(\omega),
\end{eqnarray}
\end{subequations}
where $\xi(\omega)$ is given by Eq.~(\ref{eq:xi}) and
\begin{subequations}
\label{eq:6form2}
\begin{eqnarray}
  \rho_1(\omega) &=& \rho_2(\omega) = -\frac{1}{3}{C^1 \bar \Lambda}
  \left<r^2\right> \left(\omega-1\right), \label{rho12} \\
  \rho_3(\omega) &=& \rho_4(\omega) = 0.
\end{eqnarray}
\end{subequations}
Thus, there are $1/m_Q$ corrections to $\rho_1$ and $\rho_2$ coming from phase
factors of the wave functions, together with kinetic terms. (See Appendix
\ref{app:sub:second}.)

\section{CKM matrix element}
\label{sec:Vcb}
\subsection{$\bar B \rightarrow D\ell\bar\nu$}
\label{subsec:D}
With the results given in \S\S \ref{sub:0th} and \ref{sub:1st}, we now 
calculate the CKM matrix element, $\left|V_{c\,b}\right|$. We first evaluate
the differential decay rate of the process $\bar B \rightarrow D\ell\bar\nu$ to
extract the value $\left|V_{c\,b}\right|$.
Note that since we have neglected the isospin invariance in
our formulation, there is no distinction between $\bar B_0$ and $B^-$ in the following.
Similarly, $D^{(*)}$ means either $D$ or $D^*$, which is equal to $D^0$ and $D^+$ or $D^{*0}$
and $D^{*+}$, respectively, in our formulation.
The expression for this differential decay rate is given by
\begin{eqnarray}
  {d\Gamma\over d\omega}&=&{G_F^2\over 48\pi^3}\left|V_{cb}\right|^2
  M_{D}^3\left(M_B+M_{D}\right)^2(\omega^2-1)^{3/2} {\cal F}_D(\omega)^2,
\end{eqnarray}
where we have defined
\begin{eqnarray}
  {\cal F}_D(\omega) &=&\xi_+(\omega)-\frac{1-r}{1+r} \xi_-(\omega),\quad
  r=\frac{M_D}{M_B}.
\end{eqnarray}
Using the above expressions and the form factors obtained in the previous
subsections, we can evaluate values of the form factor ${\cal F}_D(\omega)$ at
zero recoil and its first derivative.
\begin{flushleft}
i) Zeroth-order in $1/m_Q$
\end{flushleft}
In this case, since $\xi_+(\omega)=\xi_V(\omega)=\xi_{A_1}(\omega)=\xi_{A_3}
(\omega)=\xi(\omega)$ and $\xi_-(\omega)=\xi_{A_2}(\omega)=0$, given by
Eq.~(\ref{eq:6xi}), we have
\begin{equation}
  {\cal F}_D(1)=\xi(1)=1,\qquad {{\cal F}_D}'(1)=\xi'(1)=-1.44.
\end{equation}
\begin{flushleft}
ii) First-order in $1/m_Q$
\end{flushleft}
In this case, Eq.~(\ref{eq:6form1}), together with Eq.~(\ref{eq:6form2})
gives
%
%\begin{subequations}
\begin{eqnarray}
  {\cal F}_D(1)=\xi_+(1)-\frac{1-r}{1+r} \xi_-(1)=1.07, \qquad
  {{\cal F}_D}'(1)=-0.875. \label{eqD}
\end{eqnarray}
%\end{subequations}
%

%
% CLEO : VF=0.0416±0.0047±0.0037  error=0.0060
% Belle: VF=(4.11±0.44±0.52)×10-2 error=0.0068
%
The combined recent experimental data of CLEO \cite{CLEO99} and Belle
\cite{Belle02} lead to the following value of the product of the form factor and
the CKM matrix element :
\[
  {\cal F}_D(1)\left| V_{cb}\right|=0.0414\pm 0.0064.
\]
Using our prediction Eq.~(\ref{eqD}), we find the value of CKM matrix element to be
\begin{equation}
  \left| V_{cb}\right|= 0.0387\pm 0.0060.
\end{equation}

\subsection{$\bar B \rightarrow D^{*}\ell\bar\nu$}
\label{subsec:Dstar}
As noted by
Neubert \cite{Neubert}, the differential decay rate of the process
$\bar B \rightarrow D^{*}\ell\bar\nu$ is the best quantity to extract the value
$\left|V_{c\,b}\right|$. The expression for this differential decay rate
is given by
\begin{eqnarray}
  &&{d\Gamma\over d\omega}={G_F^2\over 48\pi^3}\left|V_{cb}\right|^2
  M_{D^*}^3\left(M_B-M_{D^*}\right)^2\sqrt{\omega^2-1}(\omega+1)^2 
  \nonumber \\
  &&\quad\times\,\left[1+{4\omega\over \omega+1}{1-2\omega r^*+r^{*2}\over(1-r^*)^2}
  \right]{\cal F}_{D^*}(\omega)^2,
\end{eqnarray}
where we have defined
\begin{eqnarray}
  &&(1-r^*)^2\left[1+\frac{4\omega}{\omega+1}\frac{1-2\omega r^*+r^{*2}}{(1-r^*)^2}
  \right]{\cal F}_{D^*}(\omega)^2 \nonumber \\
  &&=2(1-2\omega r^*+r^{*2})\left(\xi_{A_1}(\omega)^2+
  \frac{\omega-1}{\omega+1}\xi_V(\omega)^2\right)+
  \Big\{(\omega-r^*)\xi_{A_1}(\omega)  \nonumber \\
  &&\qquad-(\omega-1)\left[\xi_{A_3}(\omega)+r^*\xi_{A_2}(\omega)\right]\Big\}^2
  \nonumber \\
  &&=\left\{2(1-2\omega r^*+r^{*2})\left(1+\frac{\omega-1}{\omega+1}R_1(\omega)^2\right)
  +\left[(\omega-r^*)-(\omega-1)R_2(\omega)\right]^2\right\}
  \nonumber \\
  &&\quad\times\,\xi_{A_1}(\omega)^2,
\end{eqnarray}
and
\begin{equation}
  r^*=\frac{M_{D^*}}{M_B},\qquad
  R_1(\omega)=\frac{\xi_V(\omega)}{\xi_{A_1}(\omega)},\qquad
  R_2(\omega)=\frac{\xi_{A_3}(\omega)+r^*\xi_{A_2}(\omega)}{\xi_{A_1}(\omega)}.
\end{equation}
Similarly to \S \ref{subsec:D}, one can evaluate values of the form factor
${\cal F}_{D^*}(\omega)$ at zero recoil and its first derivative.
\begin{flushleft}
i) Zeroth-order in $1/m_Q$
\end{flushleft}
In this case, we have
\begin{equation}
  {\cal F}_{D^*}(1)=\xi(1)=1,\qquad {{\cal F}_{D^*}}'(1)=\xi'(1)=-1.44.
\end{equation}
\begin{flushleft}
ii) First-order in $1/m_Q$
\end{flushleft}
In this case, we have
\begin{subequations}
\label{eqDstar}
\begin{eqnarray}
  &&{\cal F}_{D^*}(1)=\xi(1)=1, \qquad  {{\cal F}_{D^*}}'(1)=-1.09, \\
  && R_1(1)=1.45, \quad R'_1(1)=-0.222, \quad R_2(1)=0.942, \quad R_2'(1)=0.0286,
\end{eqnarray}
\end{subequations}
%
% m_c=1.032, m_b=4.639, r=M_D*/M_B=2008/5279=0.380375
%
where the first equation is consistent with the so-called Luke theorem,
\cite{POL} and we have used $m_c=1.032$ GeV, $m_b=4.639$ GeV, and
$r^*=M_{D^*}/M_B=0.3804$
used in Ref. \citen{Matsuki06} and $C^1=0.19022$ in Eq.~(\ref{rho12}).
The estimated values given by Eqs.~(\ref{eqD}) and (\ref{eqDstar}) in our paper
are consistent with other theoretical values, except for $R_2'(1)$, whose values are
given in Ref. \citen{EGF06}. The value of $R_2'(1)$ is one order of magnitude smaller
than the other theoretical values, and there is yet no experimental data for it.

%
% CLEO : FV=0.0431±0.0013(stat)±0.0018(syst) error=0.0022
% Belle: FV=(3.54±0.19±0.18)×10-2            error=0.0026
% BaBar: FV=(35.5±0.3±1.6)×10-3             error=0.0016
%
The combined fit of the experimental data from CLEO \cite{CLEO032}, Belle
\cite{Belle022}, and BaBar \cite{BaBar05} gives
\[
  {\cal F}_{D^*}(1)\left| V_{cb}\right|=0.0380\pm 0.0021.
\]
Using our value of ${\cal F}_{D^*}(1)=1$, we have
\begin{equation}
  \left| V_{cb}\right|=0.0380\pm 0.0021.
\end{equation}
\begin{table}[t!]
\caption{Comparison of our model predictions for the values of the product
${\cal F}_{D^*}(1)|V_{cb}|$ with theoretical predictions and experimental data \cite{EGF06}. }
\label{FD1}
%\begin{center}
\begin{tabular}{@{\hspace{0.3cm}}c@{\hspace{0.3cm}}c@{\hspace{0.4cm}}
c@{\hspace{0.4cm}}c@{\hspace{0.4cm}}c@{\hspace{0.4cm}}c@{\hspace{0.4cm}}c@{\hspace{0.3cm}}}
\hline
\hline
param. & our & 
Ref. \citen{EGF06} \\
\hline
%$\rho^2$ 
%& 1.00 & 0.86
%& $\left\{ {\begin{array}{*{18}c}
%   {0.91(15)(6)^a}  \\
%   {1.61}  \\
%  \end{array}} \right.$
%& $0.79(6)^a$
%& $0.81(12)^a$ & $1.39(10)(33)^b$ \\
%
${\cal F}_{D^*}|V_{cb}|$ 
& 0.0380(21) & 0.0343(12) \\
%%%%%%%%%%%%%%%%%%%%%%%%%%%
\hline
\end{tabular}
%\end{center}
\end{table}
\begin{table}[t!]
\label{FD2}
%\begin{center}
\begin{tabular}{@{\hspace{0.3cm}}c@{\hspace{0.3cm}}c@{\hspace{0.3cm}}
c@{\hspace{0.3cm}}c@{\hspace{0.3cm}}c@{\hspace{0.3cm}}c@{\hspace{0.3cm}}c@{\hspace{0.3cm}}}
\hline
\hline
params. & 
CLEO \cite{CLEO032} &
BaBar \cite{BaBar05} & 
Belle \cite{Belle022} &
DELPHI \cite{DelPhi04} \\
\hline
${\cal F}_{D^*}|V_{cb}|$  & $\left\{ {\begin{array}{*{18}c}
   0.0376(3)(16)^a \\
   0.0328(5)^b  \\
  \end{array}} \right.$
& $\left\{ {\begin{array}{*{18}c}
   0.0431(13)(18)^a \\
   0.0360(20)^b \\
  \end{array}} \right.$
& $\left\{ {\begin{array}{*{18}c}
   0.0354(19)(18)^a \\
   0.0315(12)^b \\
  \end{array}} \right.$
& $0.0377(11)(19)^a$ \\
%%%%%%%%%%%%%%%%%%%%%%%%%%%
\hline
\end{tabular}
%\end{center}
~ \\
{$^a$~fit using the form factor $\xi_{A_1}$ parameterization of the model \cite{Boyd97}.} \\
{$^b$~fit using the form factor predictions of the model \cite{EGF06}.}
\end{table}
\begin{table}[t!]
%\begin{wraptable}{l}{\halftext}
\caption{Comparison of theoretical predictions of for the ratios $R_1(1)$ and
$R_2(1)$ and their derivatives, $R'_1(1)$ and $R'_2(1)$.}
\label{R1R2}
\begin{center}
\begin{tabular}{@{\hspace{0.3cm}}c@{\hspace{0.3cm}}c@{\hspace{0.3cm}}
c@{\hspace{0.3cm}}c@{\hspace{0.3cm}}c@{\hspace{0.3cm}}}
\hline
\hline
Ref. & $R_1(1)$ & 
$R_2(1)$ & 
$R'_1(1)$ & 
$R'_2(1)$ \\
\hline
our
& 1.45
& -0.22
& 0.94
& 0.029 \\
\citen{EGF06}
& 1.39
& -0.23
& 0.92
& 0.12 \\
\citen{Grinstein02}
& 1.25
& -0.10
& 0.81
& 0.11 \\
\citen{Caprini98}
& 1.27
& -0.12
& 0.80
& 0.11 \\
\citen{REVIEW,Neubert96}
& 1.35
& -0.22
& 0.79
& 0.15 \\
\citen{Melikhov00}
& 1.15
& 
& 0.94
&  \\
\citen{Albertus05}
& 1.01(2)
& 
& 1.04(1)
&  \\
%%%%%%%%%%%%%%%%%%%%%%%%%%%
\hline
\end{tabular}
\end{center}
%\end{wraptable}
\end{table}
\section{Summary and discussion}
\label{sec:summary}
In this paper, we have formulated a method for constructing Lorentz-invariant
amplitudes by examining four kinds of Lorentz-boosted systems after normalizing
their amplitudes. Although all four are not always consistent
for higher orders in $1/m_Q$, they may be corrected by comparing with 
other methods, for instance, relativistic and kinematic results.
Despite of this defect, we believe that this formulation certainly provides
a method to calculate reliable amplitudes from wave functions in the rest frame.

As an example of this formulation, we have calculated the semi-leptonic weak form
factors up to first-order in $1/m_Q$ following a previous study, \cite{Seo97}
and using the results obtained in that paper. In this paper, we have found the
following results.
\begin{enumerate}
\item The Isgur-Wise function has the following form, up to first-order in
$1/m_Q$ and in $(\omega-1)$ :
\[
  \xi(\omega)=1-\left({1\over 2}+{1\over 3}{\bar \Lambda}^2\left<r^2\right>
  \right)\left(\omega-1\right),\quad 
  {\bar \Lambda}={\bar \Lambda_u}=\lim_{m_Q\rightarrow \infty}\tilde E_D
  =\lim_{m_Q\rightarrow \infty}\tilde E_B.
\]
Hence, we have
\[
  \xi(1)=1,\quad \xi'(1)=-{1\over 2}-{1\over 3}{\bar \Lambda}^2
  \left<r^2\right>.
\]
Here, since $\bar\Lambda$ depends only on the light quark mass and we treat only the
heavy mesons $D$, $D^*$, and $B$, which include only $u$ and $d$ light quarks
with $m_u=m_d$, the subscript of $\bar\Lambda_u$ expresses this fact.
\item We find that there is no contribution to the six form factors from correction terms
at first-order in $m_Q$ for the rest-frame wave functions. That is, no terms,
except for the first in Eq.~(\ref{eq:psi01}), which include negative and positive
energy states of a heavy quark contribute to the physical quantities.
\item The first-order corrections are derived from phase factors of the wave
functions and also given by kinetic terms, and there are no contributions from
the first-order corrections to the wave functions. They are explicitly given by
Eqs.~(\ref{eq:6form1}) and (\ref{eq:6form2}). That is, in the terminology
of Neubert and Rieckert in Ref. \citen{POL}, we have
%
%\begin{subequations}
\begin{eqnarray*}
  \rho_1(\omega) &=& \rho_2(\omega) = -\frac{1}{3}{C^1 \bar \Lambda}
  \left<r^2\right> \left(\omega-1\right), \quad
  \rho_3(\omega) = \rho_4(\omega) = 0.
\end{eqnarray*}
%\end{subequations}
%
\item We have calculated the values for the form factor ${\cal F}(\omega)$ at
zero recoil and/or their first derivatives up to first-order in $1/m_Q$ as
\begin{eqnarray*}
  &&{\cal F}_D(1)=1.07, \qquad {{\cal F}_D}'(1)=-0.875, \\
  &&{\cal F}_{D^*}(1)=\xi(1)=1, \qquad  {{\cal F}_{D^*}}'(1)=-1.09, \\
  && R_1(1)=1.45, \quad R_1(1)=-0.222, \quad R_2(1)=0.942, \quad R_2'(1)=0.0286.
\end{eqnarray*}
The first equations here were obtained by analyzing the $\bar B \rightarrow D\ell\bar\nu$
process, and the second by analyzing $\bar B \rightarrow D^{*}\ell\bar\nu$. These values
are consistent with experimental data as well as other theoretical estimates listed in
Tables I and II of Ref. \citen{EGF06}.
\item The above values can be used to estimate the CKM matrix element
$\left|V_{cb}\right|$, which we have obtained as
\[
  \left| V_{cb}\right|= 0.0387\pm 0.0060
\]
for the $\bar B \rightarrow D\ell\bar\nu$ process, and
\[
  \left| V_{cb}\right|=0.0380\pm 0.0021
\]
for the $\bar B \rightarrow D^*\ell\nu$ process. These values are consistent with the value
in PDG, \cite{PDG}
\[
  \left| V_{cb}\right|=0.0409\pm 0.0018 \quad({\rm exclusive}).
\]
Here to obtain theoretical predictions for $\left| V_{cb}\right|$, we have ignored
theoretical undertainties, although experimental errors are taken into account.
\end{enumerate}

We have developed a method to obtain relativistically invariant results using
rest-frame wave functions,
and to do this, four different Lorentz-boosted frames were adopted to check the
validity of our results. The same form is obtained for the Isgur-Wise function in
all four cases up to zeroth-order in $1/m_Q$ and first-order in $(\omega-1)$.
However, the first-order corrections in $1/m_Q$ are not the same in all four
cases. Only the case denoted by 2-ii) in the main text, i.e., in the Breit 
frame with $t'=x'^0=y'^0$ both for $\bar B$ and $D^{(*)}$, gives results consistent
with the relativistic ones given in Ref. \citen{POL}.

\appendix
\section{Schr\"odinger Equation}
\label{app:FYeq}
The Schr\"odinger equation given by Eq.~(\ref{eq:FYeq}) is derived by calculating
\begin{equation}
  \left<X';P_{X'}\,\{\ell'\}\left|{\cal H}\right|X;P_X\,\{\ell\}\right>
  =\left<X;P_X\,\{\ell\}\left|{\cal P}^0\right|X;P_X\,\{\ell\}\right>,
  \label{app:eq:FYeq1}
\end{equation}
with ${{\cal P}^0}$ being the 0-th component of the four-momentum 
${{\cal P}^\mu}$, and by varying with respect to $\psi_{X'}^{\ell'\,*}$. 
Here the Hamiltonian density is given by
\begin{eqnarray}
  {\cal H}&&=\int d^3x\;\{q^{c\dagger}(\vec x)({\vec \alpha}_q\cdot {\vec p}_q 
  +\beta_q m_q)q^c(\vec x)+Q^{\dagger}(\vec x)({\vec \alpha}_Q\cdot {\vec p}_Q 
  +\beta_Q m_Q)Q(\vec x)\} \nonumber \\ 
  &&\qquad +\int d^3x d^3y\; q^{c\dagger}(\vec x)\beta_q O_{qi}q^c(\vec x)
  V_{ij}({\vec x}-{\vec y})Q^{\dagger}(\vec y)\beta_Q O_{Qj}Q(\vec y).
\end{eqnarray}
The lhs of Eq.~(\ref{app:eq:FYeq1}) can be approximated as
\begin{eqnarray}
  &&\left<X';P_{X'}\,\{\ell'\}\left|{\cal H}\right|X;P_X\,\{\ell\}\right>
  \nonumber \\
  &&=\left<X;P_X\,\{\ell\}\left|\int d^3x \int d^3y {q_\alpha^c}^\dagger (t,
  \vec x)q_\alpha^c(t,\vec x)Q_\beta^\dagger(t,\vec y)Q_\beta(t,\vec y)
  {\cal H}\right|X;P_X\,\{\ell\}\right> \nonumber \\
  &&\simeq \int d^3x \int d^3y\left<X';P_{X'}\,\{\ell'\}\left|Q_\beta^\dagger
  (t,\vec y){q_\alpha^c}^\dagger (t,\vec x)\right|0\right> \nonumber \\
  &&\quad\times\,\left<0\left|
  q_\alpha^c(t,\vec x)Q_\beta(t,\vec y){\cal H} Q_\delta^\dagger(t,\vec y)
  q_\gamma^{c \dagger}(t,\vec x) q_\gamma^c(t,\vec x) Q_\delta(t,\vec y)
  \right|X;P_X\,\{\ell\}\right> \nonumber \\
  &&\simeq \int d^3x \int d^3y {\psi_{X'}^{\ell'\,*}}_{\alpha\,\beta}
  \left(\left(0,\vec x-\vec y\right); P_{X'}\right)\,H_{\alpha\,\gamma,\,
  \beta\,\delta}{\psi_X^{\ell}}_{\gamma\,\delta}\left(\left(0,\vec x-
  \vec y\right); P_X\right) \nonumber \\
  &&\quad\times\,e^{-i\left(P_X-P_{X'}\right)\cdot y}, 
  \label{app:eq:FYeq2}
\end{eqnarray}
where the number operators for $q^c$ and $Q$ have been inserted, and the vacuum
dominance has been used among the intermediate states.
The rhs of Eq.~(\ref{app:eq:FYeq1}) is given by
\begin{equation}
  \left<X';P_{X'}\,\{\ell'\}\left|{\cal P}^0\right|X;P_X\,\{\ell\}\right>
  =\sqrt{M_X^2+{\vec P_X}^2}\left<X';P_{X'}\,\{\ell'\}\left|
  \right.X;P_X\,\{\ell\}\right>, \label{app:eq:FYeq3}
\end{equation}
and thus we obtain the Schr\"odinger equation, Eq.~(\ref{eq:FYeq}).
\section{Lorentz Boost and Normalization}
\label{app:normal}
The derivations of Eqs.~(\ref{eq:casei}), (\ref{eq:caseii}) and (\ref{eq:norm2}) are 
similar to those given in \S \ref{app:FYeq}. Setting ${\cal H}=1$ 
in Eq.~(\ref{app:eq:FYeq2}), we obtain
\begin{eqnarray}
  &&\left<X;P_{X'}\,\{\ell'\}\left|\right.X;P_X\,\{\ell\}\right>
  \nonumber \\
  &&\simeq \int d^3x \int d^3y\; 
  {\psi_{X}^{\ell'\,*}}_{\alpha\,\beta}\left(\left(0,\vec x-\vec y\right); 
  P_{X'}\right)\,{\psi_X^{\ell}}_{\alpha\,\beta}\left(\left(0,\vec x-\vec y
  \right); P_X\right)\,e^{-i\left(P_X-P_{X'}\right)\cdot y} \nonumber\\
  &&=\left(2\pi\right)^3\;\delta^3\left(\vec P_{X'}-\vec P_X\right)
  \int d^3z\; {\rm tr}\left[{\psi_{X}^{\ell'}}^\dagger\left(\left(0,
  \vec z\right);P_{X}\right)\psi_X^\ell\left(\left(0,\vec z\right);P_X\right)
  \right]. \label{app:eq:norm0}
\end{eqnarray}
The lhs is given by
\begin{equation}
  \left<X;P_X'\,\{\ell'\}\left.\right|X;P_X\,\{\ell\}\right>
  =(2\pi)^3\,2{P_X}_0\,\delta^3\left(\vec P_X-\vec P_X'\right)\delta_
  {\ell'\,\ell},\label{app:eq:norm1}
\end{equation}
and hence
\begin{equation}
  \int d^3z\;{\rm tr}\left[{\psi_{X}^{\ell'}}^\dagger\left(\left(0,
  \vec z\right);P_{X}\right)\psi_X^\ell\left(\left(0,\vec z\right);P_X\right)
  \right]\simeq 2{P_X}_0\,\delta_{\ell'\,\ell} \label{app:eq:norm2}
\end{equation}
holds.

We have to calculate the normalization of the wave function in the moving frame
in two cases, i) $t=x^0=y^0$ and ii) $t'=x'^0=y'^0$. To do so, we need to have
a relation between the rest-frame (RF) and Lorentz-boosted (LB) wave functions.
\begin{flushleft}
i) $t=x^0=y^0$ and $x'^0\ne y'^0$ (equal time in the rest frame)
\end{flushleft}
The relation between RF and LB is derived as follows.
By definition, we have
\begin{eqnarray}
  &&\left<0\left|\,q_\alpha^c(t,\vec x)Q_\beta(t,\vec y)
  \right|X;\left(M_X,\vec 0\right)\{\ell\}\right>
  \nonumber\\
  &&=G_{\alpha\,\gamma}^{-1}G_{\beta\,\delta}^{-1}\,
  \left<0\left|q_\gamma^c(x'^0,\vec x')Q_\delta(y'^0,\vec y')
  \right|X;P_X\,\{\ell\}\right>.
\end{eqnarray}
Hence, we obtain
\begin{eqnarray}
  &&\left<0\left|\,q_\alpha^c(x'^0,\vec x')Q_\beta(y'^0,\vec y')
  \right|X;P_X\,\{\ell\}\right> \nonumber\\
  &&\simeq \psi^\ell_{X\,\alpha\beta}\left((0,\vec x'-\vec y');P_X\right)\,
  \exp\left(i[-M_X t+(M_X-m_Q)\gamma^2\beta(x^3-y^3)]\right) \nonumber\\
  &&=G_{\alpha\,\gamma}G_{\beta\,\delta}\,
  \left<0\left|\,q_\gamma^c(t,\vec x)Q_\delta(t,\vec y)
  \right|X;\left(M_X,\vec 0\right)\{\ell\}\right>  \nonumber\\
  &&=G_{\alpha\,\gamma}G_{\beta\,\delta}\,\psi_{X\,\gamma \delta}^\ell
  \left(\vec x-\vec y\right) \,e^{-iM_X t},\label{app:eq:boost}
\end{eqnarray}
where use has been made of the approximation
\begin{equation}
  Q_\beta\left(y'^0,\vec y'\right)\simeq \exp\left[-im_Q\gamma\left(y'^0
  -x'^0\right)\right]Q_\beta\left(x'^0,\vec y'\right).
\end{equation}
Rewriting Eq.~(\ref{app:eq:boost}), we obtain the relation
\begin{equation}
  {\psi_X^{\ell}}_{\alpha\,\beta}\left(\left(0,\vec x\right); P_X\right)\simeq
  G_{\alpha\,\gamma}G_{\beta\,\delta}\,{\psi_X^{\ell}}_{\gamma\,\delta}\left(
  \left(0,\vec x_\perp,\gamma^{-1}x^3\right); \left(M_X,\vec 0\right)\right)\,
  e^{i\left(M_X-m_Q\right)\gamma\,\beta\,x^3}. \label{app:eq:casei}
\end{equation}
Using this equation, we find
\begin{eqnarray}
  &&\int\; d^3x\; \psi^{\ell\,*}_{X\;\alpha\beta}\left(\left(0,\vec 
  x\right);  P_X\right)\psi^\ell_{X\;\alpha\beta}\left(\left(0,\vec x\right)
  ;P_X\right)  \nonumber \\
   && =\gamma\,\int\; d^3x\;\psi^{X\;\ell\,*}_{\alpha\beta}\left(
  \vec x\right)\left(G^2\right)_{\alpha\gamma} \left(G^2\right)_{\beta\delta}
  \psi^\ell_{X\;\gamma\delta}\left(\vec x\right) \nonumber \\
  && =\gamma\int d^3x\;{\rm tr}\left[\psi^{\ell\,\dagger}_X\left(
  \vec x\right) G^2 \psi^\ell_X\left(\vec x\right) G^{2\,\rm T}\right]
  =2M\gamma^3. \label{app:eq:normP1}
\end{eqnarray}
\begin{flushleft}
ii) $t'=x'^0=y'^0$ and $x^0\ne y^0$ (equal time in the moving frame)
\end{flushleft}
Similarly to the case i), the relation between RF and LB is, in this case,
given by
\begin{equation}
  {\psi_X^{\ell}}_{\alpha\,\beta}\left(\left(0,\vec x\right); P_X\right)\simeq
  G_{\alpha\,\gamma}G_{\beta\,\delta}\,{\psi_X^{\ell}}_{\gamma\,\delta}\left(
  \left(0,\vec x_\perp,\gamma x^3\right); \left(M_X,\vec 0\right)\right)\,
  e^{i\left(M_X-m_Q\right)\gamma\,\beta\,x^3}. \label{app:eq:caseii}
\end{equation}
Then, using Eq.~(\ref{app:eq:caseii}) we obtain
\begin{eqnarray}
  &&\int\; d^3x\; \psi^{\ell\,*}_{X\;\alpha\beta}\left(\left(0,\vec x\right);
  P_X\right)\psi^\ell_{X\;\alpha\beta}\left(\left(0,\vec x\right);P_X\right) \nonumber \\
  &&=\gamma^{-1}\,\int\;d^3x\;\psi^{\ell\,*}_{X\;\alpha\beta}\left(
  \vec x\right)\left(G^2\right)_{\alpha\gamma}\left(G^2\right)_{\beta\delta}
  \psi^\ell_{X\;\gamma\delta}\left(\vec x\right) \nonumber \\
  &&=\gamma^{-1}\,\int\; d^3x\;{\rm tr}\left[\psi^{\ell\,\dagger}_X
  \left(\vec x\right) G^2 \psi^\ell_X\left(\vec x\right) G^{2\,\rm T}\right]
  =2M\gamma. \label{app:eq:normP2}
\end{eqnarray}
Here use has been made of the following :
\begin{eqnarray}
  &&{1\over 2}{\rm tr}\,\left[\left(
  \begin{array}{*{20}c} 0&u_{-1}(r) \\ 
  0&-i(\vec\sigma \cdot\vec n)v_{-1}(r) \\ \end{array}
  \right)^\dagger G^2 \left(
  \begin{array}{*{20}c} 0&u_{-1}(r) \\ 
  0&-i(\vec\sigma\cdot\vec n)v_{-1}(r) \\ \end{array}\right)
  G^{2\,\rm T}\right] 
  \nonumber \\
  ~ \nonumber \\
  &&={\gamma^2\over 2}\,{\rm tr}\,\left[\left(
  \begin{array}{*{20}c} 0&0 \\ u_{-1}(r)&i(\vec
  \sigma\cdot\vec n)v_{-1}(r) \\ \end{array} \right)\left(
  \begin{array}{*{20}c} 1&\sigma^3 \beta \\ 
  \sigma^3\beta&1 \\ \end{array} \right)\left(
  \begin{array}{*{20}c} 0&u_{-1}(r) \\ 0&-i(\vec\sigma\cdot
  \vec n)v_{-1}(r) \\ \end{array} \right)\right. \nonumber \\
  &&\quad\times\,\left.\left(
  \begin{array}{*{20}c} 1&\sigma^3 \beta \\ 
  \sigma^3\beta&1\\ \end{array} \right)\right] \nonumber \\
  ~ \nonumber \\
  &&=\gamma^2 \left(u^2_{-1}+v^2_{-1}\right). \label{app:eq:formula}
\end{eqnarray}
%

%%%%%%%%%%%%%%%%%%%%%%%%%%%%%%%%%%%%%%%%%%%%%%%%%%%%%%%%%%%%%%%%%%%%%
\section{Wave Functions}
\label{app:wf}
The wave function is generally defined as
\begin{eqnarray}
  &&\left<0\left|q_\alpha^c(t,\vec x)\,Q_\beta(t,\vec y)
  \right|X;P_X\,\{\ell\}\right> \nonumber \\
  &&=\left<0\left|q_\alpha^c
  (0,\eta(\vec x-\vec y))\,Q_\beta(0,\zeta(\vec y-\vec x))
  \right|X;P_X\,\{\ell\}\right>\,e^{-iP_X^0 t} \nonumber \\
  && \equiv{\psi_X^{\ell}}_{\alpha\,\beta}
  \left(\left(0,\vec x-\vec y\right); P_X\right)\,e^{-iP_X\cdot r},
\end{eqnarray}
with
\begin{equation}
  r^\mu=\zeta x^\mu+\eta y^\mu,\quad \zeta+\eta=1.
\end{equation}
Here, without loss of generality, we adopt
\[
  \zeta=0,\qquad \eta=1.
\]
The Foldy-Wouthuysen-Tani transformation and charge conjugation are defined as
\begin{subequations}
\label{app:eq:Hfwt}
\begin{eqnarray}
    &&U_{\rm FWT}\left( p \right)=\exp \left( {W\left( p \right)
    \,\vec \gamma_Q \cdot \vec {\hat p}} \right)=\cos W+\vec \gamma_Q\cdot
    \vec{\hat p}\,\sin W,\\
    &&\vec {\hat p}={{\vec p} \over p},\quad \tan W\left( p \right)
    ={p \over {m_Q+E}},\quad E=\sqrt{{\vec p~}^2+m_Q^2}, \\
    &&U_c=i\,\gamma^0_Q\gamma^2_Q=-U_c^{-1}.
\end{eqnarray}
\end{subequations}
Here, $\vec p=\vec p_Q$ is the initial momentum of the heavy quark, and 
$U_{\rm FWT}\left( p \right)$ operates on heavy quarks. Henceforth the color,
$N_c=3$, is ignored since the form of the wave function is the same for all
colors. The quantity $\ell$ stands for a set of quantum numbers, $j$, $m$, and $k$, and
\begin{equation}
  \Psi _{j\,m}^k(\vec x) ={1\over r}\left( {\begin{array}{*{20}c}u_k(r) \\
  -i\,v_k(r)\left(\vec n\cdot\vec \sigma\right)\\ \end{array}} \right)
  \;y_{j\,m}^k (\Omega),\label{app:eq:Psi+}
\end{equation}
where $r=\left|\vec x\right|$, $\vec n=\vec x/r$, $j$ is the total angular 
momentum of the meson, $m$ is its $z$ component, $k$ is a quantum number which 
takes only values, $k=\pm j,\;\pm(j+1)\;{\rm and}\;\ne 0$, whose operator
form is given by $\hat k=-\beta_q\left(\vec\Sigma_q\cdot\vec\ell+1\right)$. 
The scalar functions $u_k(r)$ and $v_k(r)$ are polynomials of the radial 
variable $r$ and satisfy
\begin{equation}
  \int dr\left(u_k^2(r)+v_k^2(r)\right) = 1. \label{app:eq:normUV}
\end{equation}
Here, $y_{j\,m}^k (\Omega)$ are functions of angles and spinors of the total angular 
momentum, $\vec j=\vec \ell+\vec s_q+\vec s_Q$, with $\vec\ell=-i\vec r\times
\nabla$, whose first few explicit forms are given by
\begin{eqnarray}
  &&y_{00}^{-1}={1\over \sqrt{4\pi}},
  \quad y_{1m}^{-1}={1\over \sqrt{4\pi}}\sigma^m,
  \nonumber \\
  &&y_{1m}^2={-1\over\sqrt{4\pi}}{3\over\sqrt{6}}\left(n^i n^m-{1\over 3}
  \delta^{i\,m}\right)\sigma^i. \label{app:eq:yjmk}
\end{eqnarray}
These functions satisfy the follwing relation and normalization condition :
\begin{eqnarray}
  &&y_{j\,m}^{-k}=-\left(\vec n\cdot\vec\sigma\right)y_{j\,m}^k,\\
  &&{1 \over 2}{\rm tr}\left( {\int {d\Omega
  \;{y_{j'\,m'}^{k'}}^\dagger(\Omega) \;y_{j\,m}^k}}(\Omega) \right)
  =\delta^{k\,k'}\delta _{j\,j'}\delta _{m\,m'}.\label{app:eq:normaly}
\end{eqnarray}
The relative phases among $y_{j\,m}^k$ given by Eq.~(\ref{app:eq:yjmk}) are 
determined so that they give the correct relative phases among the form factors,
which are determined by Eqs.~(\ref{eq:wff1})-(\ref{eq:wff3}). 
From this point, whenever the trace of $y_{j\,m}^k$ appears, it is understood to be 
in the sense of Eq.~(\ref{app:eq:normaly}). 

The leading-order pseudoscalar state ($0^-$) corresponds to 
$\ell=(k=-1,~j=m=0)$, and hence we have the wave function
\begin{equation}
  \psi^\ell_{X\,0}\left(\vec x\right)=
  \sqrt{M_X}\,\left(\begin{array}{*{20}c}
  0&\Psi _{0\,0}^{-1}(\vec x) \end{array} \right)=
  \sqrt{M_X\over {4\pi}}{1\over r}\,\left(\begin{array}{*{20}c} 0&u_{-1}(r)\\
  0& -i(\vec n\cdot\vec\sigma) v_{-1}(r) \\ \end{array} \right),
\end{equation}
with $X=D$ or $B$. On the other hand, the leading-order vector state ($1^-$) 
has the set of quantum numbers $\ell=(k=-1,~j=1)$ and is given by
\begin{eqnarray}
  \psi^\ell_{X\,0}\left(\vec x\right)&=&
  \sqrt{M_X}\,\left(
  \begin{array}{*{20}c} 0&\Psi_{1\,m}^{-1}(\vec x) \epsilon^m \end{array}
  \right) \nonumber \\
  &=&
  \sqrt{M_X\over {4\pi}}{1\over r}\,\left(
  \begin{array}{*{20}c} 0&u_{-1}(r)\cr 0&
  -i(\vec n\cdot\vec\sigma) v_{-1}(r)\\ \end{array}
  \right)\left(\vec\epsilon\cdot\vec \sigma\right),
\end{eqnarray}
with $\epsilon^m$ being a polarization vector ($\vec \epsilon^{~2}=-1$) and 
$X=D^*$ or $B^*$.
%

%%%%%%%%%%%%%%%%%%%%%%%%%%%%%%%%%%%%%%%%%%%%%%%%%%%%%%%%%%%%%%%%%%%%%
\section{Matrix Elements}
\label{app:elements}
The light anti-quark is treated as a spectator. That is, it does not 
interact with other particles, except for gluons represented by potentials
in our model. The heavy quark has a weak vertex and its current is 
in general given by
\begin{equation}
  j(\vec x,t)={Q_{X'}^\dagger}_\alpha(t,\vec x)\,{\cal O}_{\alpha\,\beta}
  \,{Q_X}_\beta(t,\vec x),\label{app:eq:current}
\end{equation}
where ${\cal O}$ is a $4\times 4$ matrix. By inserting the number operator
of the anti-quark and by assuming vacuum dominance among intermediate
states, we obtain the following formula for the matrix element of the above current,
i.e., Eq.~(\ref{app:eq:current}), between heavy mesons :
\begin{eqnarray}
  &&\left<X';P_{X'},\{\ell'\}\left|\,j(t,\vec x)
  \right|X;P_X\,\{\ell\}\right> \nonumber \\
  &=& {\cal O}_{\alpha\,\beta}
  \left<X';P_{X'}\,\{\ell'\}\left|\,{Q_{X'}^\dagger}_\alpha(t,\vec x)
  \,\int\,d^3y\,{q_\gamma^c}^\dagger(t,\vec y)\,q_\gamma^c(t,\vec y)
  \,{Q_X}_\beta(t,\vec x)\,\right|X;P_X\,\{\ell\}\right> \nonumber \\
  &\simeq& 
  \int\,d^3y\,{\cal O}_{\alpha\,\beta}\,\left<X';P_{X'}\,\{\ell'\}\left|\,
  {Q_{X'}^\dagger}_\alpha(t,\vec x){q_\gamma^c}^\dagger(t,\vec y)
  \right|0\right> \nonumber \\
  &&\quad\times\,\left<0\left|\,q_\gamma^c(t,\vec y)\,{Q_X}_\beta(t,\vec x)
  \,\right|X;P_X\,\{\ell\}\right> \nonumber \\
  &=&\int\,d^3y\;{\rm tr}\left[{\psi_{X'}^{\ell'}}^\dagger\left(\vec y-
  \vec x;P_{X'}\right)\,\psi_X^\ell\left(\vec y-\vec x;P_X\right)\,{\cal O}
  ^{\rm T}\right]\,e^{-i\left(P_X-P_{X'}\right)\cdot x},\label{app:eq:mat1}
\end{eqnarray}
or
\begin{equation}
  \left<X';P_{X'}\,\{\ell'\}\left|\,j(0,\vec 0)
  \right|X;P_X\,\{\ell\}\right>
  =\int\,d^3x\;{\rm tr}\left[{\psi_{X'}^{\ell'}}^\dagger\left(\vec x;P_{X'}
  \right)\,\psi_X^\ell\left(\vec x;P_X\right)\,{\cal O}^{\rm T}\right], 
  \label{app:eq:mat2}
\end{equation}
where in Eq.~(\ref{app:eq:mat1}) we have assumed vacuum dominance among
intermediate states in order to make some approximations.

\section{Semi-Leptonic Weak Form Factor}
\label{app:IW}
In this appendix, we use the Foldy-Wouthuysen-Tani transformed wave
functions, $\psi_{X\; {\rm FWT}}^\ell$, instead of $\psi_X^\ell$, although the
subscript FWT is omitted for simplicity.

\subsection{Zeroth order (Isgur-Wise function)}
\label{app:sub:first}
Here we present the derivations of the Isgur-Wise function
given in the main text in four cases.
\begin{flushleft}
1-i) $\bar B$ rest-frame and $t=x^0=y^0$ :
\end{flushleft}
\begin{eqnarray}
  \xi(\omega)&=&{2\sqrt{\omega}\over(1+\omega)\sqrt{2M_B}}{
  \int d^3x\; {\rm tr}\left[
  \psi^{\ell'\,\dagger}_{D\,0}\left(\left(0,\vec x\right);P_D\right)
  \psi^\ell_{{\bar B}\,0}\left(\left(0,\vec x\right);P\right)\right]\over 
  \sqrt{\int d^3x\;{\rm tr}\left|\psi^{\ell'}_{D\,0}\left(\left(
  0,\vec x\right);P_D\right)\right|^2}} \nonumber \\
  &=&{2\sqrt{\omega}\over(1+\omega) \sqrt{2M_B 2M_D\gamma^3}}\int\; d^3x\; 
  G_{\alpha\gamma}^*G_{\beta\delta}^* \psi^{\ell'\,*}_{D\,0\,\gamma\delta}
  \left(\left(0,\vec x_\perp,\gamma^{-1}x^3\right);\left(M_D,{\vec 0}\right)
  \right) \nonumber \\
  &&\qquad \times e^{-i(M_D-m_c)\gamma Vx^3}
  \psi^\ell_{{\bar B}\,0\,\alpha\beta}\left(\vec x\right) \nonumber \\
  &=&{2\sqrt{\omega}\over 2{\overline M}(1+\omega)\gamma^{3/2}}\int d^3x\; 
  \left\{\psi^{\ell'\,*}_{D\,0\,\alpha\beta}\left(\vec x\right)+
  (\gamma^{-1}-1)x^3{\partial\over \partial x^3}\psi^{\ell'\,*}_{D\,0\,\alpha
  \beta}\left(\vec x\right)\right\} \nonumber \\
  &&\qquad\times G_{\alpha\gamma}G_{\beta\delta}
  \psi^\ell_{{\bar B}\,0\,\gamma\delta}
  \left(\vec x\right) e^{-i(M_D-m_c)\gamma Vx^3} + \; O(\beta^4) \nonumber \\
  &=&{1\over\omega}-{1\over6}\beta^2\omega{\tilde E_D}^2\left<r^2\right>-
  {\beta^2\over2}{1\over 2{\overline M}}\int d^3x\;x^3\left({\partial\over \partial
  x^3}\psi^{\ell'\,*}_{D\,0\,\alpha\beta}\left(\vec x\right)\right)
  \psi^\ell_{B\,0\,\alpha\beta}\left(\vec x\right)+ \; O(\beta^4) \nonumber \\
  &=&{1\over\omega}-{1\over6}\beta^2\omega{\tilde E_D}^2\left<r^2\right>
  +{\beta^2\over4}+ \; O(\beta^4),
\end{eqnarray}
where we have used
\begin{eqnarray}
  &&\int d^3x\;x^3\left({\partial\over \partial x^3}\psi^{\ell'\,*}_{D\,0\,
  \alpha\beta}\left(\vec x\right)\right)
  \psi^\ell_{{\bar B}\,0\,\alpha\beta}\left(\vec x\right) \nonumber \\
  &=&{2{\overline M}\over 4\pi}\int d^3x\;{x^3\over2}{\rm tr}\;\left[{\partial
  \over\partial x^3}\,\left\{{1\over r}\,\left(
  \begin{array}{*{20}c} 0&0 \\ u_{-1}&\;i 
  ({\vec n}\cdot{\vec \sigma})\,v_{-1} \\ \end{array} \right) \right\}\,
  {1\over r}\,\left(
  \begin{array}{*{20}c} 0&u_{-1} \\ 0&-i({\vec n}\cdot {\vec\sigma})v_{-1}\\
  \end{array} \right) \right]
  \nonumber \\
  &=&{2{\overline M}\over 4\pi}\int d^3x\; {x^3\over2}{\rm tr}\;\left[\left\{
  {\partial\over
  \partial x^3}\left({u_{-1}\over r}\right)\right\}\,{u_{-1}\over r}+\left\{
  {\partial\over\partial x^3}({\vec n}\cdot{\vec\sigma})\,{v_{-1}\over r}
  \right\}\left({\vec n}\cdot{\vec\sigma}\right)\,{v_{-1}\over r}\right] 
  \nonumber \\
  &=&{2{\overline M}\over 4\pi}\int d^3x\; {x^3\over4}{\partial\over\partial x^3}
  {\rm tr}\;\left[\left({u_{-1}\over r}\right)^2+({\vec n}\cdot
  {\vec\sigma})^2\left({v_{-1}\over r}\right)^2\right]=-{\overline M}\int_0^\infty
  dr\;\left(u_{-1}^2+ v_{-1}^2\right) \nonumber \\
  &=&-{\overline M},
\end{eqnarray}
with ${\overline M}=\sqrt{M_B M_D}$.

\begin{flushleft}
1-ii) $\bar B$ rest-frame and $t'=x'^0=y'^0$ :
\end{flushleft}
\begin{eqnarray}
  \xi(\omega)&=&{2\sqrt{\omega}\over(1+\omega) \sqrt{2M_B 2M_D\gamma}}\int\; 
  d^3x\; G_{\alpha\gamma}^*G_{\beta\delta}^*\psi^{\ell'\,*}_{D\,0\,\gamma
  \delta}\left(\left(0,\vec x_\perp,\gamma x^3\right);\left(M_D,{\vec 0}\right)
  \right) \nonumber \\
  &&\qquad \psi^\ell_{{\bar B}\,0\,\alpha\beta}\left(\vec x\right)
  \times e^{-i(M_D-m_c)\gamma \beta x^3} \nonumber \\
  &=&{2\sqrt{\omega}\over 2{\overline M}(1+\omega)\gamma^{1/2}}\int d^3x\;
  \left\{\psi^{\ell'\,*}_{D\,0\,\alpha\beta}\left(\vec x\right)+(\gamma-1)\,x^3
  {\partial\over \partial x^3}\psi^{\ell'\,*}_{D\,0\,\alpha\beta}
  \left(\vec x\right)\right\} \nonumber \\
  &&\qquad\times G_{\alpha\gamma}G_{\beta\delta}
  \psi^\ell_{{\bar B}\,0\,\gamma\delta}
  \left(\vec x\right) e^{-i(M_D-m_c)\gamma \beta x^3}+\;O(\beta^4) \nonumber \\
  &=&1-{1\over6}\beta^2\omega{\tilde E_D}^2\left<r^2\right>+{\beta^2\over2}
  {1\over 2{\overline M}}\int d^3x\; x^3
  \left({\partial\over \partial x^3}\psi^{\ell'\,*}_
  {D\,0\,\alpha\beta}\left(\vec x\right)\right)
  \psi^\ell_{{\bar B}\,0\,\alpha\beta}
  \left(\vec x\right)+ \; O(\beta^4) \nonumber \\
  &=&1-{1\over6}\beta^2\omega{\tilde E_D}^2\left<r^2\right>-{\beta^2\over4}
  +\;O(\beta^4).
\end{eqnarray}
\begin{flushleft}
2-i) Breit frame and $t=x^0=y^0$ :
\end{flushleft}
\begin{eqnarray}
  \xi(\omega)&=&{\int d^3x\;{\rm tr}\left[\psi^{\ell'\,\dagger}_{D\,0}
  \left(\left(0,\vec x\right);P_D\right)\psi^\ell_{B\,0}\left(\left(0,\vec x
  \right);P_B\right)\right]\over \sqrt{\int d^3x\; {\rm tr}\left|\psi^
  {\ell'}_{D\,0}\left(\left(0,\vec x\right);P_D\right)\right|^2\,
  \int d^3x\; {\rm tr}\left|\psi^\ell_{B\,0}\left(\left(0,\vec x\right);P_B
  \right)\right|^2}} \nonumber \\
  &=&{1\over \sqrt{2M_B\gamma^3 2M_D\gamma^3}}\int\; d^3x\;  \nonumber \\
  &&\quad\times\, G_{\alpha\gamma}^* G_{\beta\delta}^*
  \psi^{\ell'\,*}_{D\,0\,\gamma\delta}\left(\left(0,\vec 
  x_\perp,\gamma^{-1}x^3\right);\left(M_D,{\vec 0}\right)\right)
  e^{-i(M_D-m_c)\gamma {\beta\over 2}x^3} \nonumber \\
  &&\quad \times\, G^{-1}_{\alpha\epsilon}G^{-1}_{\beta\tau}
  \psi^\ell_{{\bar B}\,0\,
  \epsilon\tau}\left(\left(0,\vec x_\perp,\gamma^{-1}x^3\right);\left(M_B,
  {\vec 0}\right)\right) e^{-i(M_B-m_b)\gamma {\beta\over 2}x^3} \nonumber \\
  &=&{\gamma\over 2{\overline M}\gamma^3}
  \int d^3z\;\psi^{\ell'\,*}_{D\,0\,\alpha \beta}\left(\vec z\right)
  \psi^\ell_{{\bar B}\,0\,\alpha\beta}\left(\vec z\right)
  e^{-i({\tilde E_B}+{\tilde E_D})\gamma^2 {\beta\over 2}z^3} \nonumber \\
  &=&\gamma^{-2}-{1\over6}\left({\beta\over2}\right)^2\gamma^2({\tilde E_B}+{\tilde E_D})^2
  \left<r^2\right>+O(\beta^4).
\end{eqnarray}
\begin{flushleft}
2-ii) Breit frame and $t'=x'^0=y'^0$ :
\end{flushleft}
\begin{eqnarray}
  \xi(\omega)&=&{1\over \sqrt{2M_B\gamma 2M_D\gamma}}\int\; d^3x\; G_{\alpha
  \gamma}^*G_{\beta\delta}^*\psi^{\ell'\,*}_{D\,0\,\gamma\delta}\left(\left(
  0,\vec x_\perp,\gamma^{-1}x^3\right);\left(M_D,{\vec 0}\right)\right) 
  \nonumber \\
  &&\qquad\times G^{-1}_{\alpha\epsilon}G^{-1}_{\beta\tau}
  \psi^\ell_{{\bar B}\,0\,
  \epsilon\tau}\left(\left(0,\vec x_\perp,\gamma^{-1}x^3\right);\left(M_B,
  {\vec 0}\right)\right)
  e^{-i({\tilde E_B}+{\tilde E_D})\gamma {\beta\over 2}x^3} \nonumber \\
  &=&{\gamma^{-1}\over 2{\overline M}\gamma}\int d^3z\; \psi^{\ell'\,*}_{D\,0\,
  \alpha\beta}\left(\vec z\right)
  \psi^\ell_{{\bar B}\,0\,\alpha\beta}\left(\vec z
  \right)e^{-i({\tilde E_B}+{\tilde E_D}){\beta\over 2}z^3} \nonumber \\
  &=&\gamma^{-2}-{1\over6}\left({\beta\over2}\right)^2\gamma^{-2}({\tilde E_B}+{\tilde E_D}
  )^2\left<r^2\right>+O(\beta^4).
\end{eqnarray}
%
%All these four exressions reduce to the same form given by Eq.(\ref{eq:xi})
%when taking the limit of $\omega\rightarrow 1$ and up to the order of
%$(1/m_Q)^0$.

%
\subsection{First order}
\label{app:sub:second}
In this appendix, we show how to calculate the first-order
corrections to the form factors by using Eqs.~(\ref{eq:matrixFWT})-(\ref{eq:psi01})
and (\ref{eq:normP2}) and the equations in Appendix \ref{app:wf}.
Only the case 2-ii) gives six form factors that are consistent with those of
Neubert and Rieckert \cite{POL} among the four different Lorentz frames considered
in the previous subsection :
\begin{eqnarray}
  \xi_+(\omega)&=&{1\over 2\overline M\gamma^2}\int d^3z\; {\rm tr}
  \left({\psi_D^\ell}^\dagger(\vec z)\psi_{\bar B}^\ell(\vec z)
  U_B^{-1}G^{-1}G^*U_D\right)
  \,e^{-2i{\Lambda}{\beta\over 2}z^3} \nonumber \\
  &=&{1\over 2\overline M\gamma^2}\int d^3z\; {\rm tr}\left({\psi_{D\,0}^\ell}^
  \dagger(\vec z)\psi_{\bar B\,0}^\ell(\vec z)\left(1+
  {1\over2}\left[{1\over m_c}\vec\gamma\cdot\vec p_c-{1\over m_b}
  \vec\gamma\cdot\vec p_b\right]\right)\right)
  \,e^{-2i{\Lambda}{\beta\over 2}z^3} \nonumber \\
  &=& 1-\left(\frac{1}{2}+\frac{1}{3}\Lambda^2\left<r^2\right>\right)(\omega-1)
  \nonumber \\
  &=& 1-\left[\frac{1}{2}+\frac{1}{3}\left(\bar\Lambda+C^1
  \left(\frac{1}{m_c}+\frac{1}{m_b}\right)^2 + \cdots\right)
  \left<r^2\right>\right](\omega-1) \nonumber \\
  &\sim &
  \left[1-\frac{2}{3}{\bar \Lambda}C^1
  \left<r^2\right> \left(\omega-1\right)
  \left(\frac{1}{m_c}+\frac{1}{m_b}\right) \right]
  \left[1-\left({1\over 2}+{1\over 3}{\bar \Lambda}^2
  \right) \left<r^2\right> \left(\omega-1\right) \right]
  \nonumber \\
  &=&
  \left[1-\frac{2}{3}{\bar \Lambda}C^1
  \left<r^2\right> \left(\omega-1\right)
  \left(\frac{1}{m_c}+\frac{1}{m_b}\right) \right]
  \xi(\omega),
\end{eqnarray}
where $U_X$ for $X=\bar B$ and/or $D$ are defined by Eqs.~(\ref{eq:FWTpsi}) and 
(\ref{app:eq:Hfwt}) and $\overline M=\sqrt{M_BM_D}$. Here, $\Lambda$ is defined by
\begin{eqnarray}
  \Lambda &=& \frac{\tilde E_D+\tilde E_{\bar B}}{2}=
  \frac{(E_D-m_c)+(E_{\bar B}-m_b)}{2} \nonumber \\
  &=& C^0+
  \frac{C^1}{2}\left(\frac{1}{m_c}+\frac{1}{m_b}\right)+O\left(1/(m_Q)^2\right),
\end{eqnarray}
which comes from the expansion of the phase factor
$\exp\left(-2i{\Lambda}{\beta\over 2}z^3\right)$.
Here, we have $C^0=\bar \Lambda=\bar \Lambda_u$ and $C^1$ is defined in Eq.(\ref{C_X})
and depends only on $m_q=m_u=m_d$. Comparing the result with Eq.~(\ref{eq:6form1a}),
we obtain
\begin{eqnarray}
  \xi(\omega) &=& 1-\left({1\over 2}+{1\over 3}{\bar \Lambda}^2
  \right) \left<r^2\right> \left(\omega-1\right), \\
  \xi_+(\omega) &=&
  \left[1+\left(\frac{1}{m_c}+\frac{1}{m_b}\right)\rho_1(\omega) \right]
  \xi(\omega),
  \quad \rho_1(\omega) = -\frac{1}{3}{\bar \Lambda}C^1
  \left<r^2\right> \left(\omega-1\right).
\end{eqnarray}
Similarly, the five other form factors and $\rho_i(\omega)$ are given by the following :
\begin{eqnarray}
  \xi_-(\omega)&=&{1\over 2\overline M\gamma^2{\beta\over 2}}\int d^3z\; {\rm tr}
  \left({\psi_D^\ell}^\dagger(\vec z)\psi_{\bar B}^\ell(\vec z)
  U_B^{-1}G^{-1}{\alpha^3}^{\rm T}G^*U_D\right)
  \,e^{-2i{\Lambda}{\beta\over 2}z^3} \nonumber \\
  &=&-{1\over 2\overline M\gamma^2}\int d^3z\; {\rm tr}\left({\psi_{D\,0}^\ell}^
  \dagger(\vec z)\psi_{\bar B\,0}^\ell(\vec z)\left(\alpha^3+
  {1\over2}\left[{1\over m_c} p_c^3+{1\over m_b} p_b^3\right]
  \gamma^0\right)\right)
  \,e^{-2i{\Lambda}{\beta\over 2}z^3} \nonumber \\
  &=&\left(-{\bar\Lambda\over 2}+\rho_4(\omega)\right)
  \left({1\over m_c}-{1\over m_b}\right)\xi(\omega), \\
  \rho_4(\omega) &=& C^1 \left(-\frac{1}{4} +\frac{1}{6}{\bar \Lambda}^2
  \left<r^2\right> \left(\omega-1\right) \right)
  \left(\frac{1}{m_c}+\frac{1}{m_b}\right) =0+ O(1/m_Q), \\
  \xi_V(\omega)&=&{1\over 2\overline M^*\gamma^2{\beta\over 2}(-i\epsilon^{2\,*})}
  \int d^3z\; {\rm tr}
  \left({\psi_{D^*}^\ell}^\dagger(\vec z)\psi_{\bar B}^\ell(\vec z)
  U_B^{-1}G^{-1}{\alpha^1}^{\rm T}G^*U_{D^*}\right)
  \,e^{-2i{\Lambda'}{\beta\over 2}z^3} \nonumber \\
  &=&-{1\over 2\overline M^*\gamma{\beta\over 2}(-i\epsilon^{2\,*})}\int d^3z\; 
  {\rm tr}\left(
  {\psi_{D^*\,0}^\ell}^\dagger(\vec z)\psi_{\bar B\,0}^\ell(\vec z) \right.
  \nonumber \\
  &&\times \left.\left(\alpha^1+i{\beta\over2}\Sigma^2+{i\over2}\left[{1\over m_c} p_c^3-
  {1\over m_b} p_b^3\right]\gamma^0\Sigma^2\right)\right)
  e^{-2i{\Lambda'}{\beta\over 2}z^3} \nonumber \\
  &=&\left[1+{\Lambda'\over 2}\left({1\over m_c}+{1\over m_b}
    \right)\right]\xi(\omega) \nonumber \\
  &=& 
  \left[1+\frac{\bar\Lambda}{2}\left(\frac{1}{m_c}+\frac{1}{m_b}
    \right)+\left(\frac{1}{m_c}+\frac{1}{m_b}\right)\rho_1(\omega)
    \right]\xi(\omega),
  \\
  \rho_2(\omega) &=& \rho_1(\omega), \qquad \rho_4(\omega) = 0,
\end{eqnarray}
where $\vec\Sigma=\left(
\begin{array}{*{20}c} \vec\sigma&0 \\ 0&\vec\sigma \end{array} \right)$,
$\vec\epsilon$ is a static polarization vector for $D^*$, and 
$\overline M^*=\sqrt{M_BM_{D^*}}$\,.
Here, we define
\begin{eqnarray}
  \Lambda' &=& \frac{\tilde E_{D^*}+\tilde E_{\bar B}}{2}=
  \frac{(E_{D^*}-m_c)+(E_{\bar B}-m_b)}{2} \nonumber \\
  &=&
  C^0+\frac{{C^1}'}{2m_c}+\frac{C^1}{2m_b}+O\left(1/(m_Q)^2\right),
\end{eqnarray}
with ${C^0}'=C^0=\bar\Lambda=\bar\Lambda_u$,
\begin{eqnarray}
  \xi_{A_1}(\omega)&=&{1\over 2\overline M^*\gamma^2\epsilon^{1\,*}}\int d^3z\; 
  {\rm tr}\left({\psi_{D^*}^\ell}^\dagger(\vec z)\psi_{\bar B}^\ell(\vec z)
  U_B^{-1}G^{-1}\gamma_5^{\rm T}\alpha^{1\,\rm T}G^*U_D\right)
  \,e^{-2i{\Lambda'}{\beta\over 2}z^3} \nonumber \\
  &=&{1\over 2\overline M^*\gamma^2\epsilon^{1\,*}}\int d^3z\; {\rm tr}\left(
  {\psi_{D^*\,0}^\ell}^\dagger(\vec z)\psi_{\bar B\,0}^\ell(\vec z)\right. \nonumber \\
  &&\times \left.\left(\alpha^1+i{\beta\over2}\Sigma^2+{1\over2}\left[{1\over m_c} p_c^3-
  {1\over m_b} p_b^3 \right]\gamma^0\Sigma^1\right)\right)
  e^{-2i{\Lambda'}{\beta\over 2}z^3} \nonumber \\
  &=&\left[1+\frac{\Lambda'}{2}\frac{\omega-1}{\omega+1}
    \left({1\over m_c}+{1\over m_b}\right)\right]\xi(\omega) \nonumber \\
  &=&\left[1+\frac{\bar\Lambda}{2}\frac{\omega-1}{\omega+1}
    \left(\frac{1}{m_c}+\frac{1}{m_b}\right)+
    \left(\frac{1}{m_c}+\frac{1}{m_b}\right)\rho_2(\omega)
    \right]\xi(\omega),
  \\
  \rho_2(\omega) &=& \rho_1(\omega), \qquad \rho_4(\omega) = 0,
\end{eqnarray}
\begin{eqnarray}
  &&-\xi_{A_1}(\omega)+\xi_{A_2}(\omega)+\xi_{A_3}(\omega) \nonumber \\
  &&={1\over 2\overline M^*
  \gamma^4{\beta\over 2}\epsilon^{3\,*}}\int d^3z\; {\rm tr}
  \left({\psi_{D^*}^\ell}^\dagger(\vec z)\psi_{\bar B}^\ell(\vec z)
  U_B^{-1}G^{-1}\gamma_5^{\rm T}G^*U_{D^*}\right)
  \,e^{-2i{\Lambda'}{\beta\over 2}z^3} \nonumber \\
  &&=-{1\over 2\overline M^*\gamma^4{\beta\over 2}\epsilon^{3\,*}}\int d^3z\; 
  {\rm tr}\left({\psi_{D^*\,0}^\ell}^\dagger(\vec z)\psi_{\bar B\,0}^\ell
  (\vec z) \gamma_5\left(1+{1\over 2}\left[{1\over m_c} p_c^3+
  {1\over m_b} p_b^3\right]\right)\right) 
  e^{-2i{\Lambda'}{\beta\over 2}z^3} \nonumber \\
  &&={\Lambda'\over 2}\left({1\over m_c}-{1\over m_b}\right)\xi(\omega), \\
%\end{eqnarray}
%
%\begin{eqnarray}
  &&\xi_{A_1}(\omega)+{\omega-1\over \omega+1}\left(\xi_{A_2}(\omega)-
  \xi_{A_3}(\omega)\right) \nonumber \\
  &&={1\over 2\overline M^* \gamma^4\epsilon^{3\,*}}
  \int d^3z\; {\rm tr} \left({\psi_{D^*}^\ell}^\dagger(\vec z)\psi_{\bar B}^
  \ell(\vec z) U_B^{-1}G^{-1}\gamma_5^{\rm T}\alpha^{3\,\rm T}G^*U_{D^*}
  \right)%% \nonumber \\
%  &&\times 
  e^{-2i{\Lambda'}{\beta\over 2}z^3} \nonumber \\
  &&=-{1\over 2\overline M^*\gamma\epsilon^{3\,*}}\int d^3z\; 
  {\rm tr}\left({\psi_{D^*\,0}^\ell}^\dagger(\vec z)\psi_{\bar B\,0}^\ell
  (\vec z) \left(\Sigma^3-{1\over 2}\left[{1\over m_c} p_c^3-
  {1\over m_b} p_b^3\right]\gamma_5\gamma^0\right)\right) 
  e^{-2i{\Lambda'}{\beta\over 2}z^3} \nonumber \\
  &&=\left(1-{\omega-1\over \omega+1}\right)\xi(\omega).
\end{eqnarray}
From these, we derive
\begin{eqnarray}
  \xi_{A_2}(\omega)&=&
  -\frac{\bar\Lambda}{\omega+1}{1\over m_c}\xi(\omega),
  \quad
  \rho_3(\omega) = \rho_4(\omega) = 0, \\
  \xi_{A_3}(\omega)&=&
  \left[1+\frac{\bar\Lambda}{2}\left(\frac{\omega-1} 
  {\omega+1}\frac{1}{m_c}+\frac{1}{m_b}\right) +
  \left(\frac{1}{m_c}+\frac{1}{m_b}\right)\rho_1(\omega)
  \right]\xi_+(\omega),
  \\
  \rho_2(\omega) &=& \rho_1(\omega), \qquad \rho_3(\omega) =\rho_4(\omega) = 0.
\end{eqnarray}
Therefore, all the expressions for $\rho_i(\omega)$ included in $\xi_i(\omega)$ are
consistent up to first order in $1/m_Q$.
Even though the first-order corrections to the
wave functions are included
as given by Eq.~(\ref{eq:psi01}), their contributions vanish,
because of the matrix structure after taking the trace over indices. From the above
equations, we can easily reproduce our results given by Eq.~(\ref{eq:6form1}),
together with Eq.(\ref{eq:6form2}).
%
%%%%%%%%%%%%%%%%%% reference %%%%%%%%%%%%%%%%%%%
\def\Journal#1#2#3#4{{#1} {\bf #2} (#4), #3}
\def\NIM{Nucl. Instrum. Methods}
\def\NIMA{Nucl. Instrum. Methods A}
\def\NPB{Nucl. Phys. B}
\def\NPA{Nucl. Phys. A}
\def\PLB{Phys. Lett. B}
\def\PRL{Phys. Rev. Lett.}
\def\PTP{Prog. Theor. Phys.}
\def\PRD{Phys. Rev. D}
\def\ZPC{Z. Phys. C}
\def\EPJ{Eur. Phys. J. C}
\def\EPJA{Eur. Phys. J. A}
\def\PR{Phys. Rept.}
\def\IJM{Int. J. Mod. Phys. A}
\def\AP{Ann. Phys.}
\def\RMPA{Rev. Mod. Phys.}
\def\MPLA{Mod. Phys. Lett. A}

\end{document}